\documentclass[fleqn,usenatbib]{mnras}

\usepackage{newtxtext,newtxmath}

\usepackage[T1]{fontenc}

\usepackage{tipa}

\DeclareRobustCommand{\VAN}[3]{#2}
\let\VANthebibliography\thebibliography
\def\thebibliography{\DeclareRobustCommand{\VAN}[3]{##3}\VANthebibliography}

\usepackage{graphicx}	
\usepackage{amsmath}	
\usepackage{academicons} 
\usepackage{xcolor}
\usepackage{widetext}
\usepackage{url}
\usepackage[utf8]{inputenc}
\usepackage{hyperref}
\usepackage{orcidlink}
\usepackage{multicol}
\usepackage{pdflscape}

\newcommand{\kms}{\mbox{km s$^{-1}~$}} 
\newcommand{\kmse}{\mbox{km s$^{-1}$}} 
 
\newcommand{\mse}{\mbox{m s$^{-1}$}}

\newcommand{\hi}{H{\footnotesize I} }

\newcommand{\teffe}{$\rm T_{eff}$}

\newcommand{\logge}{$\log{g}$}

\newcommand{\masyre}{mas yr$^{-1}$}
\newcommand{\pmd}{$D_{\mu}$~}
\newcommand{\pmde}{$D_{\mu}$}
\newcommand{\tpm}{$\mu$~}
\newcommand{\tpme}{$\mu$}

\title[SMC Distance Bimodality]{Exploring the Origin of the Distance Bimodality of Stars in the Periphery of the Small Magellanic Cloud with APOGEE and Gaia}

\author[Almeida et al.]{Andres Almeida\orcidlink{0009-0000-0733-2479}$^{1}$\thanks{E-mail: tac6na@virginia.edu},
Steven R. Majewski\orcidlink{0000-0003-2025-3147}$^{1}$,
David L. Nidever\orcidlink{0000-0002-1793-3689}$^{2}$,
Knut A.G. Olsen\orcidlink{0000-0002-7134-8296}$^{3}$,
\newauthor
Antonela Monachesi\orcidlink{0000-0003-2325-9616}$^{4,5}$,
Nitya Kallivayalil\orcidlink{0000-0002-3204-1742}$^{1}$,
Sten Hasselquist\orcidlink{0000-0001-5388-0994}$^{6,7}$,
Yumi Choi\orcidlink{0000-0003-1680-1884}$^{3,8}$,
\newauthor
Joshua T. Povick\orcidlink{0000-0002-6553-7082}$^2$,
John C. Wilson\orcidlink{0000-0001-7828-7257}$^1$,
Doug Geisler\orcidlink{0000-0002-3900-8208}$^{4, 5, 9}$,
Richard R. Lane\orcidlink{0000-0003-1805-0316}$^{12}$,
\newauthor
Christian Nitschelm\orcidlink{0000-0003-4752-4365}$^{13}$, Jennifer S. Sobeck\orcidlink{0000-0002-4989-0353}$^{14}$
and Guy S. Stringfellow\orcidlink{0000-0003-1479-3059}$^{15}$\\
$^{1}$Department of Astronomy, University of Virginia, P.O. Box 400325, Charlottesville, VA, 22904, USA\\
$^{2}$Department of Physics, Montana State University, P.O. Box 173840, Bozeman, MT 59717-3840\\
$^{3}$NSF’s National Optical-Infrared Astronomy Research Laboratory, 950 N. Cherry Ave., Tucson, AZ 85719, USA \\
$^{4}$Instituto Multidisciplinario de Investigacion y Postgrado, Universidad de La Serena, Raul Bitran 1305, La Serena, Chile \\
$^{5}$Departamento de Astronomia, Universidad de La Serena, Av. Cisternas 1200, La Serena, Chile \\
$^{6}$Space Telescope Science Institute, 3700 San Martin Drive, Baltimore, MD 21218 \\
$^{7}$Department of Physics \& Astronomy, University of Utah, Salt Lake City, UT 84112, USA \\
$^{8}$Department of Astronomy, University of California Berkeley, Berkeley, CA 94720, USA \\
$^{9}$Departmento de Astronom\'{i}a, Universidad de Concepci\'{o}n, Casilla 160-C Concepci\'{o}n, Chile \\
$^{10}$ Instituto de Investigaci\'{o}n Multidisciplinario en Ciencia y Tecnolog\'{i}a, Universidad de La
Serena. Avenida Ra\'{u}l Bitr\'{a}n S/N, La Serena, Chile\\
$^{11}$Departamento de Astronom\'{i}a, Facultad de Ciencias, Universidad de La Serena. Av.
Juan Cisternas 1200, La Serena, Chile \\
$^{12}$Centro de Investigaci\'{o}n en Astronom\'{i}a, Universidad Bernardo O'Higgins, Avenida Viel 1497, Santiago, Chile \\
$^{13}$Centro de Astronom{\'i}a (CITEVA), Universidad de Antofagasta, Avenida Angamos 601, Antofagasta 1270300, Chile \\
$^{14}$Department of Astronomy, University of Washington, Seattle, WA 98195-1700\\
$^{15}$Center for Astrophysics and Space Astronomy, Department of Astrophysical and Planetary Sciences, University of Colorado, Boulder, CO, 80309-0389, USA
}

\date{Accepted XXX. Received YYY; in original form ZZZ}

\pubyear{2023}

\begin{document}
\label{firstpage}
\pagerange{\pageref{firstpage}--\pageref{lastpage}}
\maketitle

\begin{abstract}

The Magellanic Cloud system represents a unique laboratory for study of 
both interacting dwarf galaxies and
the ongoing process of the formation of the Milky Way and its halo.
We focus on one aspect of this complex, 3-body interaction --- the dynamical perturbation of the Small Magellanic Cloud (SMC) by the Large Magellanic Cloud (LMC), and specifically potential tidal effects 
on the SMC's eastern side.
Using 
{\it Gaia} astrometry and the precise radial velocities and multi-element chemical abundances from APOGEE-2 DR17, 
we explore the well-known distance bimodality on the eastern side of the SMC.  Through 
estimated stellar distances, proper motions, and radial velocities, 
we characterize the kinematics of the two populations in the bimodality and compare their properties with those of SMC populations elsewhere.
Moreover, while all regions explored by APOGEE seem to show a single chemical enrichment history, the metallicity distribution function (MDF), 
of the ``far'' 
stars on the eastern periphery of the SMC is found to resemble that
for the more metal-poor fields of the western periphery, whereas the MDF 
for the ``near'' stars on the eastern periphery resembles that 
for stars in the SMC center.
The closer eastern periphery stars also show radial velocities (corrected for SMC rotation and bulk motion) that are, on average, approaching us relative to all other SMC populations sampled. 
We interpret these trends as evidence that the near stars on the eastern side of the SMC represent material pulled out of the central SMC as part of its tidal interaction with the LMC.
\end{abstract}


\section{Introduction}
\label{sec:intro}

\begin{figure*}
    \centering
    \includegraphics[scale=0.26]{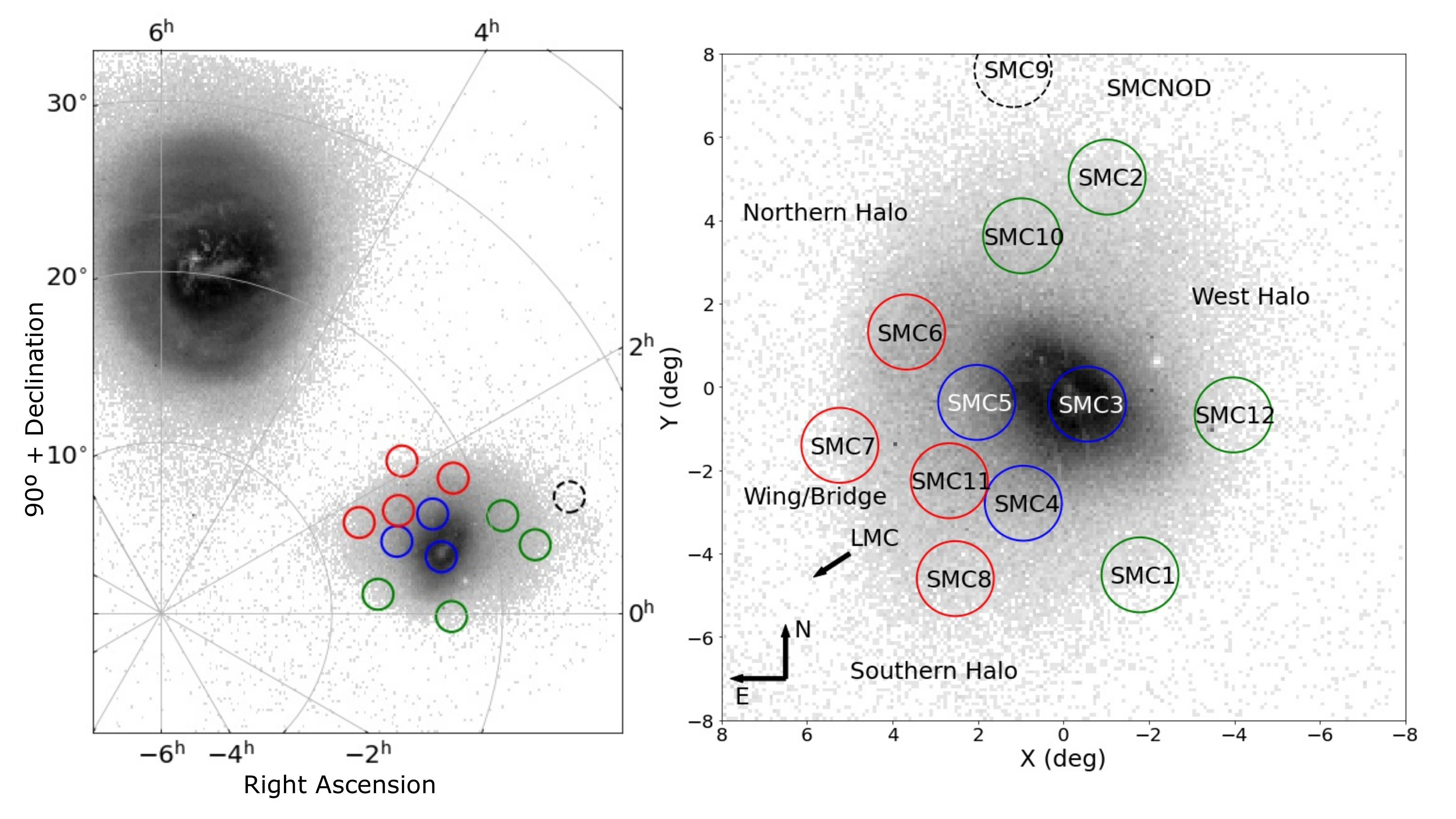}
    \caption{\label{footprint} ({\it Left}) 
    The RGB stellar density of the LMC and SMC using {\it Gaia} DR3 data in a polar projection of celestial coordinates. Overplotted on the SMC are the APOGEE fields to give a notion of their distance and orientation with respect to the LMC. ({\it Right}) The SMC RGB star density map shown in SMC-centered celestial coordinates with APOGEE fields representing the East region (red circles), the Center region (blue circles), and the West region (green circles),
   based on the scoring system described in the Appendix. The arrow indicates the direction towards the center of the LMC. 
   }
\label{fig:fields}
\end{figure*}

It is widely accepted that the Magellanic Clouds (MCs) provide critical, proximate laboratories for the study of satellite systems, late infall, minor mergers, dwarf irregular galaxies, and the interaction of such systems with one another.  Numerous studies over the past decades have exploited these unique prototypes 
for such investigations, but the scale of this attention is mushrooming with the advent of large systematic astrometric, photometric, and spectroscopic surveys \citep{VMC2011,DES2016,Smash2017,Delve2021}.  These large observational databases have led to the identification of numerous, hitherto unknown gaseous \citep{Putman2003} and stellar substructures around each the Large Magellanic Cloud \citep[LMC; e.g.,][]{Choi2018a,Belokurov2019,Nidever2019,Gaia2021,Cullinane2022a,Cullinane2022b}  and the Small Magellanic Cloud \citep[SMC; e.g.,][]{Pieres2017,Belokurov2019,Massana2020,Gaia2021}.  In the case of the diffuse peripheral SMC features (see Fig.~\ref{fig:fields}), some of that diffuse stellar material may be associated with the Magellanic Bridge, a feature likely to be of tidal origin, with stars being pulled out of the SMC \citep[e.g.,][]{Putman2003,Besla2012,Nidever2013,Zivick2019} during their infall into the Milky Way \citep{Besla2007,Kallivayalil2013}. Although a portion of the gas present in some structures can be also explained by ram pressure from the Milky Way \citep[]{Tatton2021}. These gaseous structures are easily detected in \hi maps and comprise the Leading Arm (LA) and the Magellanic Stream (MS). There is another structure of gas and stars that connect the LMC and SMC called the Magellanic Bridge (MB), which also has a strong \hi signature \citep[e.g.,][]{Putman2003,Nidever2008}.
Other observed properties of the SMC itself have also been interpreted within the context of its tidal disruption.  For example, on the eastern side of the SMC --- the side closer to the LMC --- the distribution of red clump (RC) stars has been found
to exhibit a distance bimodality, with the two populations separated by $\sim$10 kpc \citep{Hatzidimitriou1989,Nidever2013,Subramanian2017,Tatton2021,ElYoussoufi2021,James2021}.  
The two RC populations are found to be distinct in their radial velocity (RV) and proper motion distributions \cite{Omkumar2021}.  This spatial bimodality and its interpretation is mimicked by kinematical studies of red giant branch (RGB) stars in the eastern SMC periphery, where their radial velocities (RVs) also reveal a bimodal distribution, with main peaks separated by $\sim$35--45 \kms \citep{James2021}, and for which significant differences in the proper motions are seen with respect to fields on the western side of the SMC  \citep{Zivick2018,Omkumar2021,James2021}.

The origin of this bimodal distribution of distances, proper motions, and radial velocities has been proposed to be as a result of past interaction of the LMC with the SMC, where the foreground structure is postulated to be tracing a tidal extension of the latter galaxy \citep{Nidever2013}. Based on the simulations of the Magellanic System \citep{Diaz2012}, it has been proposed that the formation of the stellar substructures along with the gaseous features of the MCs have a tidal origin from the past interaction of the Clouds, creating a foreground extension of material torn from the disk of the SMC.



Most of the above studies of the SMC bimodality were limited to a radius of $\le4\deg$. Furthermore, none of these studies included an analysis of the chemistry of the foreground and background populations in the bimodality.  However, because of the known gradients in radial metallicity within the SMC \citep[][Povick et al., in prep.]{Dobbie2014,munoz2023}, it should be possible to verify whether the hypothesis that its disk is the source of the foreground material in the eastern side bimodality by comparing the chemistry of the foreground material with that of stars in other parts of the SMC, including the disk and farther out.


Here we undertake just such a study.  We exploit data obtained by the Sloan Digital Sky Survey IV's (SDSS-IV's) APOGEE-2 survey (Majewski et al., in prep.), combined with astrometry from {\it Gaia} DR3 \citep{Gaia2021}, not only to verify the previously reported kinematical differences in the near and far populations in the eastern side of the SMC, but to explore their chemical attributes as well.
A particular focus of the Southern Hemisphere component of the APOGEE-2 survey was to obtain significant coverage of the Magellanic Clouds \citep[e.g.,][]{Nidever2020}.  APOGEE's high resolution, near-infrared, multifiber spectroscopy yields very accurate \citep[$\sim$0.1 \kmse,][]{Nidever2015} RVs for SMC RGB stars as well as unprecedented insights into the multi-element chemistry of stars across the face of the SMC. These APOGEE data provide new ways to investigate the origins of the various spatio-kinematical anomalies previously reported for stars in the ``bimodality region'' on the eastern side of the SMC, and over a larger angular extent (to a radius $\sim 6\deg$) than previously explored.

While our use of red giant branch stars means that distances are less accurate than in the case of the studies using RC standard candles, we nevertheless can make reliable assertions about the bimodality.  For example, we find that the {\it Gaia} DR3 proper motions show a different behaviour between the western and eastern sides of the SMC, with an extra component only present in the latter, a result consistent with previous studies that used {\it Gaia} DR2 \citep{Omkumar2021}.
We also demonstrate that the metallicity distribution function (MDF), [$\alpha$/Fe]-[Fe/H], and detailed chemical abundance patterns of the farther eastern SMC stars resemble those same distributions for complementary stars on the western SMC periphery, while the MDF, [$\alpha$/Fe]-[Fe/H], and chemical distributions of the closer eastern SMC stars resemble those distributions of the more metal-rich SMC center.  Thus, apart from our verification of previously reported trends in proper motion and radial velocity in the bimodality regions, the main result of our investigation is that chemical analysis affirms that while all of the APOGEE stars share an SMC chemistry, the foreground population of the bimodality is more closely linked to the inner SMC region than to populations of stars farther out in the SMC periphery.  This suggests a dynamical link between the foreground eastern side population and the central regions/disk population of the SMC.  


The layout of this contribution is as follows: In \S \ref{sec:data}, we describe in more detail the data sets exploited in our analysis, which are based on the
SDSS-IV Data Release 17 (DR17) and {\it Gaia} DR3 \citep{Gaia2021}.  This includes a discussion of how we deal with distances for our red giant branch star sample (\S \ref{subsec:distances}).
We also describe how we break up the APOGEE fields into three main groups --- East, Center, and West --- for our analyses (\S \ref{subsec:smcregions}).
In \S \ref{sec:starproperties}, we 
describe various SMC stellar properties we have measured, such as distance, radial velocity, total proper motion and metallicity, and explore whether and how the character of these three regions differ, and to see whether and how the eastern bimodality manifests itself in our data.
In \S \ref{sec:CloserLook}, we explore more closely the SMC's eastern side, in particular, to define the characteristics of the near and far populations found there.
Finally, in \S \ref{sec:discussion}, we summarize our main results and conclusions.


\section{Data}
\label{sec:data}

\subsection{APOGEE and {\it Gaia} Catalogs}
\label{sec:APOGEE-Gaia}

The Apache Point Observatory Galactic Evolution Experiment (APOGEE, \citealt{Majewski2017}), part of the Sloan Digital Sky Survey (SDSS) in its phases SDSS-III (Eisenstein et al. 2011) and SDSS-IV (\citealt{Blanton2017}), is an infrared spectroscopic survey of stars sampling all Galactic stellar populations, from the inner bulge, throughout the disk, and in the Milky Way halo.  Originally, APOGEE, as with all previous SDSS projects, operated only in the Northern Hemisphere using the 2.5-m Sloan Telescope at Apache Point \citep{Gunn2006}. However, after installation of a second APOGEE spectrograph (\citealt{Wilson2019}) on the 2.5-m duPont telescope at Las Campanas Observatory \citep{bv73}, the ``APOGEE-2" project in SDSS-IV also procured a high-resolution ($R \sim 22,500$), $H$-band spectra in the Southern Hemisphere, including for thousands of stars (predominantly on the red and asymptotic giant branches) sampling in and around the Large and Small Magellanic Clouds.  Indeed, surveying the Clouds was a primary motivation for extending the APOGEE project to the Southern Hemisphere.

The APOGEE infrared spectra are reduced and analyzed to produce stellar atmospheric parameters (\teffe, \logge, [M/H], [$\alpha$/Fe], etc.) and abundances for multiple chemical elements using the APOGEE reduction pipeline \citep{dln15} and the APOGEE Stellar Parameters and Chemical Abundance Pipeline (ASPCAP; \citealt{Garcia2016}). The latter is based on the \texttt{FERRE}\footnote{\url{https://github.com/callendeprieto/ferre}} code written by \citet{AllendePrieto2006}, and obtains stellar atmospheric parameters by finding the best match in a library of synthetic spectra.
We use the specific data products coming from SDSS Data Release 17 (DR17; Holtzman et al., in prep.). These include metallicities and [$\alpha$/Fe]
derived from the APOGEE spectra using a grid of MARCS stellar atmospheres \citep{Gustafsson2008,Jonsson2020}, and an $H$-band line list updated from \citet{Smith2021}, which itselfis updated \citep{shetrone2015} to include lines for the s-process elements Ce and Nd \citep{Cunha2017,Hasselquist2016}. The grid of synthetic spectra \citep{zamora2015} is generated using the \texttt{Synspec} code \citep{Hubeny2011}, which enables nLTE calculations for the elements Na, Mg, Ca, and K from \citet{Osorio2020}.

The APOGEE reduction pipeline used for DR17 includes a new code for measuring heliocentric radial velocities ($V_{\rm helio}$) 
called {\texttt{Doppler}} \citep{Doppler2021}\footnote{\url{https://github.com/dnidever/doppler}}. This is particularly relevant here, because the {\texttt{Doppler}} algorithm was fine-tuned specifically to improve the derivation of radial velocities for faint sources having many visits, as is the case for the SMC stars, by forward-modeling all of the visit spectra simultaneously with a consistent spectral model.
 
Targeting for SDSS-III/APOGEE is described in \citet{zas13}, while that for SDSS-IV/APOGEE-2 survey is described in \citet{zas17}, \citet{Beaton2021}, and \citet{Santana2021}.
Here we make use of data from the specific collection of stars that were targeted as part of the APOGEE-2 Magellanic Clouds key project.
The selection of APOGEE Magellanic Cloud fields and specific stars in those fields for this key project are described in more detail in \citet{2020ApJ...895...88N}.\footnote{We do not include SMC Field 9 in our analysis due to the low number ($\sim$~3) of RGB stars measured in that APOGEE field.}
The location and angular span of the specific APOGEE-2 SMC fields analyzed here are shown in Figure \ref{fig:fields}, while data for the fields is given in Table \ref{tab:table_fields}, which includes for each field, respectively, the name of the field, central position, the field ``group'' (``West'', ``Center'', ``East'' --- see \S\ref{subsec:smcregions}),
the radial distance from the SMC center, the number of red giant stars included in our analysis, the mean RV and dispersion for each field, and the mean metallicity, [Fe/H].
The collection of fields includes one sampling of the SMC center (``SMC3'') as well as others distributed out to its nominal tidal radius in multiple directions.

\begin{table*}
    \label{tab:table_fields}
    \caption{Information about the SMC APOGEE Fields}
\begin{tabular}{cccccccccc}

    \hline
    \hline
    Name & RA & DEC. & Region & R$_{\rm SMC}$ & N$_{\rm RGB}$ & $\langle \rm RV \rangle$ & $\langle \sigma_{\rm RV} \rangle$ & $\langle$[Fe/H]$\rangle$ & $\langle\sigma_{[\rm Fe/H]}\rangle$\\
         & h:m:s & \degr:\arcmin:\arcsec & & \degr & & \kmse & \kmse & & \\
    \hline
    SMC1 & 00:20:16 &  $-$77:13:22 & West & 4.86 &  33  &  160.0  &  22.0  & $-$1.10 & 0.23 \\
    SMC2 & 00:41:58 &  $-$67:45:25 & West &  5.15 &  88  &  127.7  & 18.7 & $-$1.28 & 0.28 \\
    SMC10 & 01:03:49 & $-$69:10:04 & West &  3.77 &  92  &  135.2 & 22.2 &  $-$1.18 & 0.24\\
    SMC12 & 23:57:50 & $-$73:02:16 & West &  4.02 &  24  &  143.5 &  16.1 &  $-$1.15 & 0.22\\
    &  &  &   &   &   &   & & \\
    SMC3 & 00:45:00 &  $-$73:13:44 & Center &  0.69 &  183  &  145.1 &  23.5  &  $-$0.90 & 0.18 \\
    SMC4 & 01:07:56 &  $-$75:35:34 & Center &  2.95 &  164  &  168.8  & 22.6  & $-$1.14 & 0.24 \\
    SMC5 & 01:20:41 &  $-$73:04:48 & Center &  2.06 &  118  &  155.7 & 24.1 & $-$0.99 & 0.17 \\
    &  &  &   &   &   &   &  &\\
    SMC6 & 01:38:29 &  $-$71:09:10 & East &  3.91 &  171  &  138.2 & 21.2 &  $-$1.08 & 0.26 \\
    SMC7 & 02:07:23 &  $-$73:24:28 & East &  5.43 &  39  &  150.2 &  14.0  &  $-$1.21 & 0.34  \\
    SMC8 & 01:38:41 &  $-$77:11:13 & East &  5.26 &  29  &  156.3 &  23.7 &  $-$1.14 & 0.31 \\
    SMC11 & 01:33:41 & $-$74:48:21 & East &  3.47 &  128  &  159.2 & 21.5 &  $-$1.09 & 0.23\\
    \hline
\end{tabular}
\end{table*}

We combine the APOGEE-2 catalog of SMC stars with information from {\it Gaia} Data Release 3 (DR3).  The latter provides proper motions and trigonometric parallaxes for each APOGEE SMC star.  While the parallaxes are not useful for deriving distances for MC stars directly (\S \ref{subsec:distances}), they can be used to identify foreground stars (e.g., nearby dwarf stars misidentified as distant giant stars). On the other hand, the proper motions are useful indicators of the stellar transverse motions and, therefore, play a critical role in our analysis. For selecting MC members and to avoid Milky Way contamination, we use a combination of ASPCAP stellar parameters and RVs along with proper motions from {\it Gaia} DR3. From the APOGEE-2 catalog, we selected stars with \teffe$<5200$ K and \logge$ < 3.4$ and remove stars with radial velocities greater than 220 \kms and lower than 80 \kmse.
We also only consider stars with APOGEE spectra having $S/N > 40$.
In terms of the {\it Gaia} parameters, we followed the same proper motion selection criteria as in \citet{2020ApJ...895...88N}. 
In the end, the total number of RGB stars that are adopted as members of the SMC is 1,069.

 

\begin{figure}
    \centering
    \includegraphics[scale=0.35]{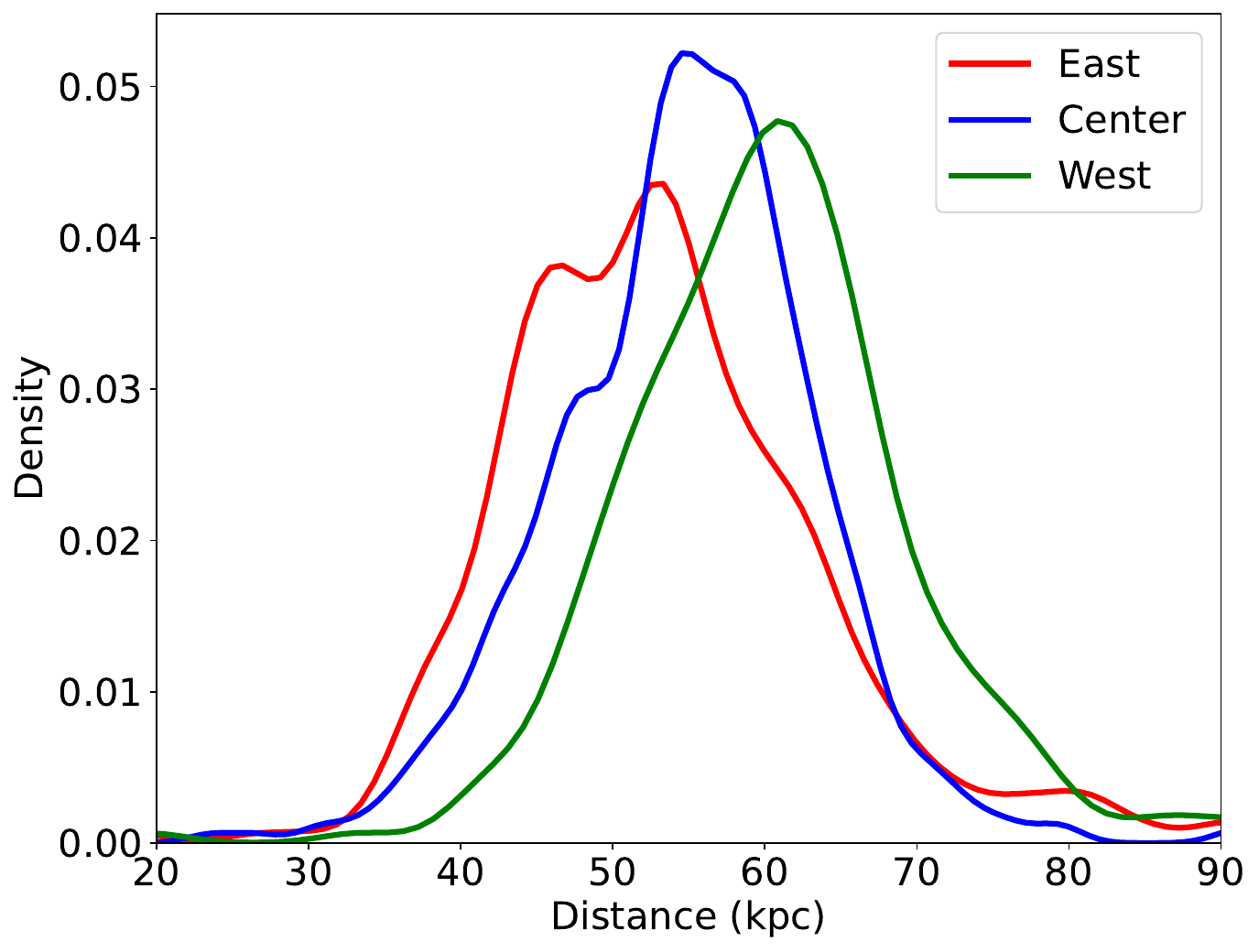}
    \caption{\label{dist-hist3} The density distribution of stellar distances in the East, Center and West regions using NMSU distance estimates. }
   \label{fig:Dist_dist}
\end{figure}

\begin{figure}
    \centering
    \includegraphics[scale=0.35]{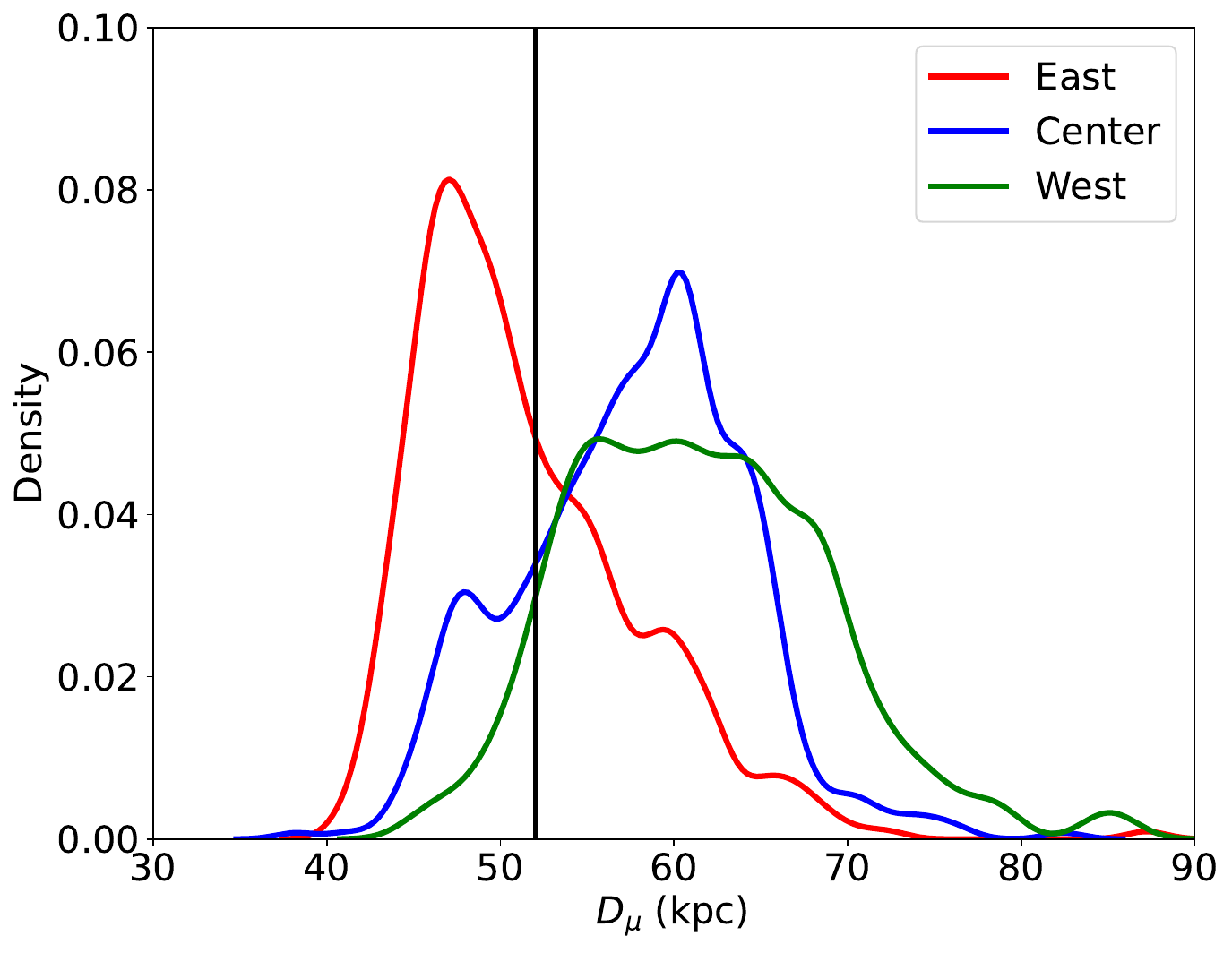}
    \caption{\label{AllPMD} 
    The proper motion distance density distribution of the three primary regions of the SMC. The Eastern region shows a large population of stars with values smaller than $\sim$52 kpc (the vertical line) and peaked at $\sim$45 kpc that is not prevalent in the West or Center regions, and which skews the overall distribution of the East group closer.
    The Center and West distributions peak at $\sim$60 kpc.}

\end{figure}

\subsection{Estimating Distances}
\label{subsec:distancesproxy}
 
Distances are challenging to derive for stars at the distances of the MCs, which are beyond the range of simple trigonometric parallaxes for single stars using {\it Gaia} data.  Several efforts have been made to exploit the stellar atmospheric parameters derived from APOGEE spectra to estimate distances.  We investigated three of the resulting distance catalogs: astroNN (\citealt{2019MNRAS.489.2079L}), StarHorse (\citealt{2020A&A...638A..76Q}), and NMSU (Holtzman et al., in prep.).  In the end, we chose to use NMSU distances because StarHorse showed a peculiar systematic distance error in a significant fraction of the APOGEE SMC stars while astroNN gave an unrealistic median distance of 40 kpc for the APOGEE SMC stars.
Nevertheless, in all three catalogs that were tested, distances seemed to be underestimated.
To compensate for this, we re-calibrated the distances using literature distances for globular clusters (GCs) and MW dwarf galaxies to yield a mean distance closer to the more recent systemic SMC distance of $\sim$60 kpc  \citep{deGrijs2015}.  The equation used for correcting the distance modulus is:
\begin{equation}
DM_{\text{corr}}=-1.3854+1.1034\times DM_{\text{NMSU}}
\end{equation}

However, for some of our analyses where we wish to have more precise {\em relative} distances of stars than offered by even the NMSU distances, we can use a proxy for distances based on stellar proper motions.  The rationale for this ``proper motion distance'' (\pmde) is that the systemic space motion of the SMC system is much larger than the internal variations of parts within it; thus, variations in observed proper motion for stars in the SMC will be dominated by the variation in distance.  Moreover, the {\it Gaia} proper motions are much better measured than any SMC star distances gauged in other ways. Therefore, we define \pmd as
\begin{equation}
D_{\mu} = v_{tan} / (4.74 ~ \mu)
\end{equation}
where \tpm is the total {\it Gaia} proper motion (in \masyre) of the star, $v_{tan}$ is the systemic SMC tangential velocity (adopted here as 398 \kmse), and \pmd is in kpc. 
Using this metric, the center of the SMC, with a mean \tpm = 1.40 \masyre, is 60 kpc. 
We find that $D_{\mu}$ as a distance proxy provides more consistent relative distances with less scatter given that much of the transverse motion of the SMC stars comes from the shared bulk motion of the SMC.\footnote{A similar methodology has been employed for these same SMC fields in Povick et al. (in prep.), but on a field by field basis.}  (In this way, $D_{\mu}$ works much like the traditional ``reduced proper motion'' methodology that has a long history as a tool in Galactic stellar population studies --- e.g., \citealt{Luyten1922}).


 

\subsection{Definition of SMC Regions}
\label{subsec:smcregions}

A primary goal of this paper is to investigate the properties of the two populations in the distance bimodality
\citep{Hatzidimitriou1989,Nidever2013} in the SMC periphery on its side that  faces the LMC (East).  To determine whether the properties of the stellar populations in this apparently tidally perturbed region are, in fact, different or more complex than other regions as a result of the perturbation, it is important to define the APOGEE fields that represent this eastern region as well as those that define appropriate control regions, which should be fields at a similar radius from the SMC center, but in unperturbed regions of the SMC periphery.  In addition, because it is possible that the perturbation may involve stars pulled out from the center of the SMC, another useful comparison sample is stars from the central parts of the SMC.

We attempted to assign APOGEE fields to the most important SMC regions for our purposes (i.e., the region already known to be unusual by its distance bimodality) in a logical way. We attempted to do so in an impartial, quantitative way using shared characteristics. Because we want to explore chemistry in particular, we want to avoid that as a criterion.  Therefore, a scoring system (described in the Appendix) based on joint kinematics (i.e., radial velocities and proper motions) and distance was used to guide our definition of the fields most uniformly representing the side of the SMC that also shows
the distance bimodality, hereafter referred to as the ``East fields''. From our scoring analysis, the Eastern group is represented by fields SMC 6, 7, 8, and 11, which not only share similar distance and kinematical properties (see Appendix), but also happen to be those closest to the LMC and oriented towards the Magellanic Bridge (although this proximity to the LMC was not used as part of the scoring system).  The remaining SMC periphery fields at the same angular separation from the SMC center but at other positions away from the LMC --- fields SMC 1, 2, 10, and 12, called the ``West fields'' --- then constitute a control sample for the East group.
Finally, the remaining fields --- fields SMC 3, 4 and 5 --- are close to one another and situated around the center of the SMC and sample predominantly its inner population.  We can readily discriminate the fields that belong in the East region from those in the Center fields based on the strongly differentiated distance and kinematical properties between them
(see Table \ref{tab:spec_sum}1).  We note that had we simply sorted fields based on spatial position on the sky, we would have naturally come to the same basic organization into three groups.\footnote{Field SMC4, while perhaps intermediary in position between the Center and East regions, has individual stellar targets that are heavily waited towards the SMC center (see Fig.~\ref{Mapbinned3} below), and so is more logically a center field.}

In the following sections, we use the above division of APOGEE SMC fields to understand similarities and differences of characteristics (kinematics and chemistry) of stars in the East fields, and the two apparent populations along this line of sight, compared to those of stars in the Center and West groupings.

\begin{figure*}
    \centering
    \includegraphics[scale=0.27]{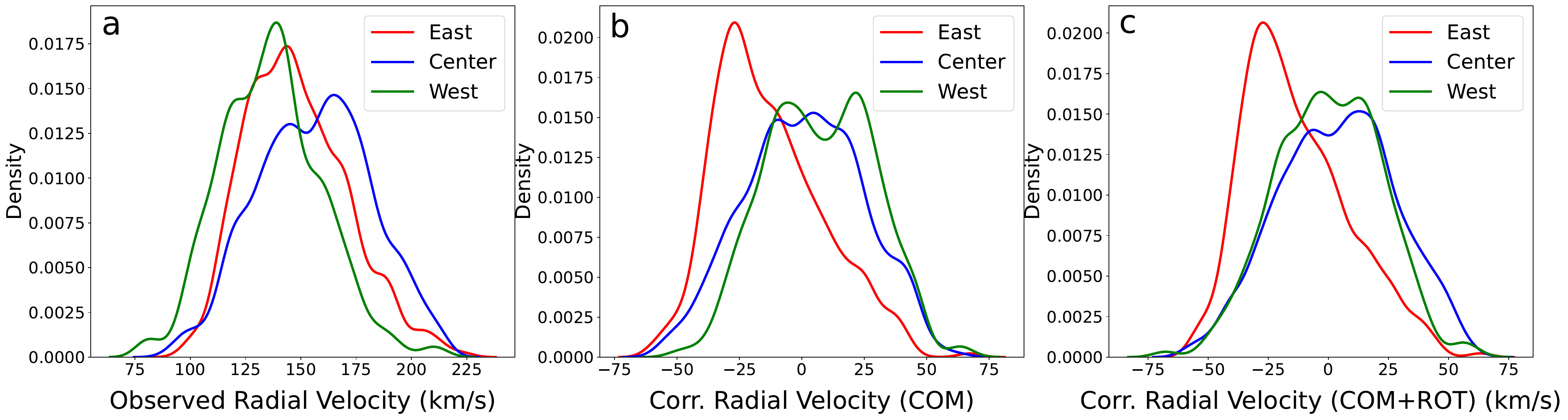}
    \caption{\label{AllVhelio} 
    The radial velocity distribution of the three primary SMC regions. Panel (a) shows the density distribution of the radial velocities as measured directly by APOGEE. Panel (b) presents the same radial velocities, accounting for SMC center of mass (COM) movement and in that reference frame. Panel (b) shows clearly that stars in the Center and West regions share a similar velocity distribution, whereas a large portion of the stars in the East region are moving away from the SMC (and toward us). Panel (c) presents the radial velocities corrected for both the SMC center of mass movement and rotation (ROT). Panel (c) looks very similar to panel (b) except for an even closer match of the Center and West distributions.} 
\end{figure*}


\section{Properties of the APOGEE SMC Stars}
\label{sec:starproperties}

Before proceeding to a detailed exploration of the previously reported bimodality on the eastern side of the SMC, in this section we first revisit the question of whether, how, and where the APOGEE fields across the SMC show distinct properties from one another, with an eye towards 
how 
the bimodality might show up in the properties of stars sampled by APOGEE on the SMC's eastern side.


\subsection{Distances}
\label{subsec:distances}

Figure \ref{dist-hist3} shows the distance distribution of the three SMC regions, initially adopting the distance estimates from the NMSU APOGEE distance catalog. As discussed in \S\ref{sec:intro}, previous studies using red clump stars found that there is a larger line-of-sight depth on the Eastern side of the SMC, particularly 
in the MB region, which shows a bimodal distance distribution \citep{Nidever2013,Subramanian2017,Tatton2021,ElYoussoufi2021,James2021}.
Our data, based on the NMSU distances for RGB stars, show the Center and West regions of the SMC to have a clear peak at $\sim$60 kpc, which can be identified as the systemic distance of the SMC. 
Meanwhile, the East region shows this same peak along with potentially a second one at $\sim$45 kpc, 
although given that distance estimates for the RGB stars carry high uncertainties, we refrain from identifying our Eastern region distance distribution as truly ``bimodal'', at least for now. 
Nevertheless, our distance distribution seems consistent both with the presence of a distance bimodality (albeit poorly distinguished with our distances) previously observed on the SMC's eastern side as well as an overall shift in the mean distances of the RGB stars on the eastern side, as previously found by \citet{Groenewegen2019} using SMC red giants.

The significantly different distance distributions of the East region from the other regions is also born out by the proper motion proxy for distance that we defined in \S\ref{subsec:distancesproxy}.  Figure~\ref{AllPMD} shows the \pmd distribution for the East region to be heavily skewed to closer distances, with a peak at $\sim$45 kpc, whereas the Center and West regions clearly resemble one another in being heavily populated around a peak at $\sim$60 kpc.

In conclusion, while we do not see a strong bimodality in our distance data, the following analysis of our much more precise kinematical and chemical properties of SMC stars make evident that the broadened distance distribution we do observe on the eastern side of the SMC likely results from the superposition along the line-of-sight of two groups of stars with different chemodynamical properties.

\begin{figure*}
    \centering
    \includegraphics[scale=0.3]{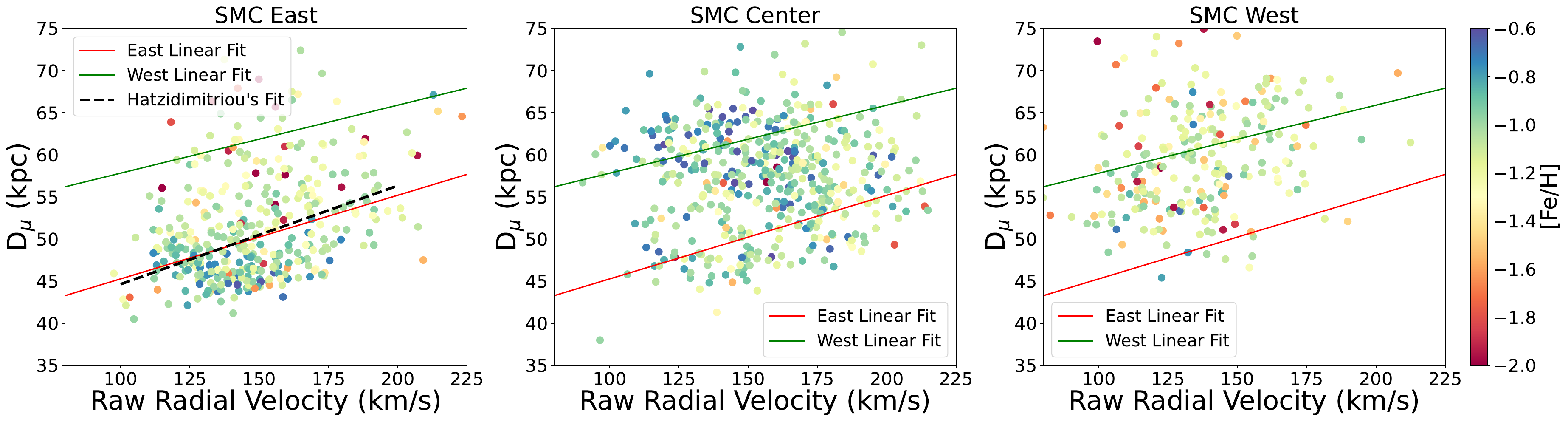}
    \caption{\label{AllVhelioDist}  The trend of $D_{\mu}$ 
with raw heliocentric radial velocity for our three primary regions. The points are color-coded by their metallicity.
The colored lines are linear fits to the trends for the East ({\it red}) and West ({\it green}) fields.  
The dashed line is the trend found on the Eastern side of the SMC by \citet{1993MNRAS.261..873H}, which matches our own results extremely well.
}
\end{figure*}


\subsection{Radial Velocities}
\label{subsec:rvs}

To investigate the kinematics of the three main SMC regions, 
we first use the heliocentric radial velocities (RVs) 
provided by the APOGEE spectral reduction pipeline, which have uncertainties of around $\sim$100 \mse. Figure \ref{AllVhelio}a shows the radial velocity distribution of stars in the three primary SMC regions under study. This RV distribution is similar in character to that found by 
\citet{James2021}, 
but it is challenging to compare directly to their results because our data cover a smaller total area, spread out in a more sparse fashion, and reach farther away from the center of the SMC. 
In our data, the East region clearly shows an RV distribution with one peak centered at about 140 \kms and asymmetric tails, with the high-RV tail extending longer to higher RVs than the low-RV tail extends to lower velocities.
In contrast, the Center fields, while showing a very similar low-RV distribution as the East region, including a ``peak'' at about 140 \kmse, also seem to have superimposed a slightly more dominant, high-RV population with a peak at about 170 \kmse.  (A small representation by stars in this higher RV population seen in the Center fields may account for the high-RV tail of the RV distribution in the East region.) The broad, perhaps ``double-peaked'' RV distribution of the Center region suggests a bimodality in this parameter for the Center fields.



However, the directly measured RV distribution is difficult to interpret because it is modulated by SMC rotation as well as the bulk motion of the SMC and the varying projection of that motion onto RVs over the large angular extent of the SMC.  
Figure \ref{AllVhelio}c shows the RV distribution corrected for SMC rotation and bulk motion using the model by \citep{Zivick2021}. This model, summarized in their Table 2, determined best-fit parameters describing the dynamical center, center of mass motion, inclination, line of nodes position angle, rotational velocity, and tidal expansion rate using a combination of {\it Gaia} DR2 proper motion measurements of SMC red giants and the line-of-sight velocity measurement of the SMC center of mass from \citet{DeLeo2020}; the SMC distance used in their analysis was that measured by \citet{Jacyszyn2020} from RR Lyrae stars.  For this work, we computed the terms describing the contribution of the center of mass bulk motion and the internal rotation in the line of sight direction at the positions of our stars using the formalism developed by \citet{vdm2002}, and subtracted these terms from our measured RVs.
In this corrected frame of reference, the velocity distribution in the Center and the West are almost identical to each other, while the one in the East is rather different, showing a clear peak at around $-$35 \kmse, a feature not present in the other regions. A large fraction of the stars in the East are moving toward us, likely pulled out of the SMC in the last tidal interaction with the LMC about 150--200 Myr \citep{Zivick2018}. This kinematical difference with the rest of the stars in the SMC may explain the origin of the bimodal distance distribution previously identified in this region (\S\ref{subsec:eastrvs}).

At first glance, the raw RV distribution (Fig.~\ref{AllVhelio}a) of the West region looks very similar to that of the East region, suggesting that somehow they may share the same kinematics.  However, after correction for bulk motion and SMC rotation (Fig.~\ref{AllVhelio}c) we find 
the East fields to have a quite distinct RV distribution from both the West and the Center fields.
Figure \ref{AllVhelioDist} shows the $D_{\mu}$ distance versus raw RV trends for the three regions in our study, and with stars color-coded by their [Fe/H].
This figure was inspired by the work of \citet{1993MNRAS.261..873H}, who found a trend of distance with the RVs for SMC regions closest to the LMC. 
However, as can be seen in Figure \ref{AllVhelioDist}, such trends are seen in all three of the East, Center, and West regions, although the trend is shallower and offset in distance for the West fields.
Figure \ref{AllVhelioDist} shows how the East and Center field distributions differ from the West fields in having a group of stars receding from the SMC towards us.


\subsection{Proper Motions of SMC Stars}
\label{subsec:propermotions}

The {\it Gaia} proper motions of the APOGEE SMC stars offer an additional, two-dimensional, kinematical signature to inform our analysis.

\begin{figure*}
    \centering
    \includegraphics[scale=0.295]{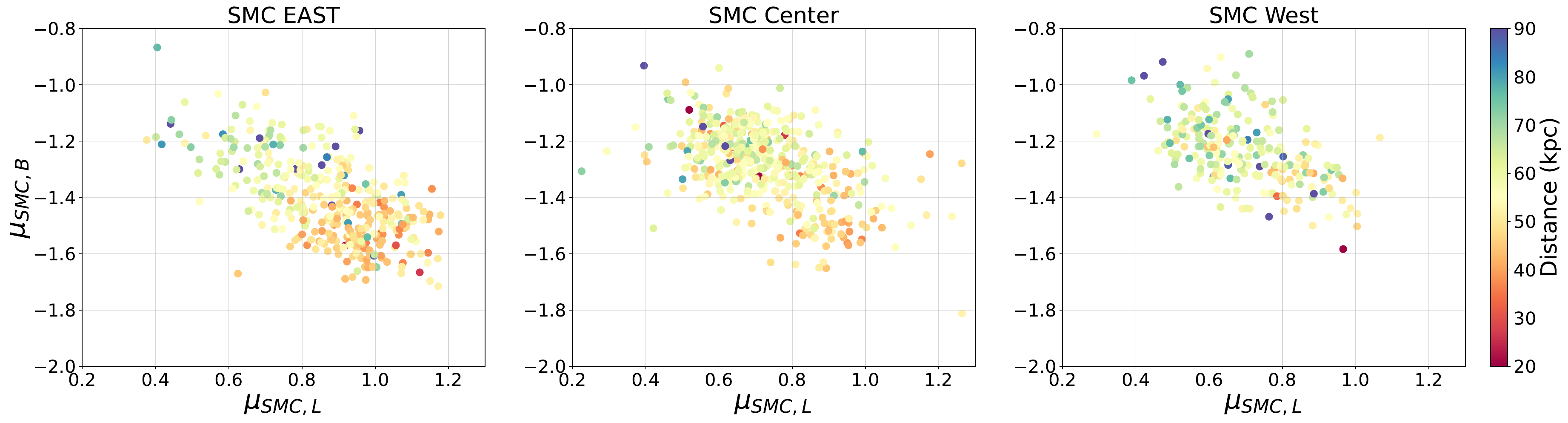}
    \caption{\label{AllPM-MS} Proper motion vector point diagrams of the three regions, based on {\it Gaia} DR3 data and with points color-coded by their NMSU distance. It is noteworthy
    how the center of the proper motion distribution shifts from the center to the upper left
    as we progress from the East to Center and then West regions.  The color-coding shows that these distributions are actually comprised of stars at two rather distinct distances, with the stars in the lower right proper motion ``lobe'' seen predominantly in the East region (but with a smaller representation in the Center region) being some 10-20 kpc closer than the stars in the upper left proper motion ``lobe''. 
    }
\end{figure*}

\begin{figure*}
    \centering
    \includegraphics[scale=0.35]{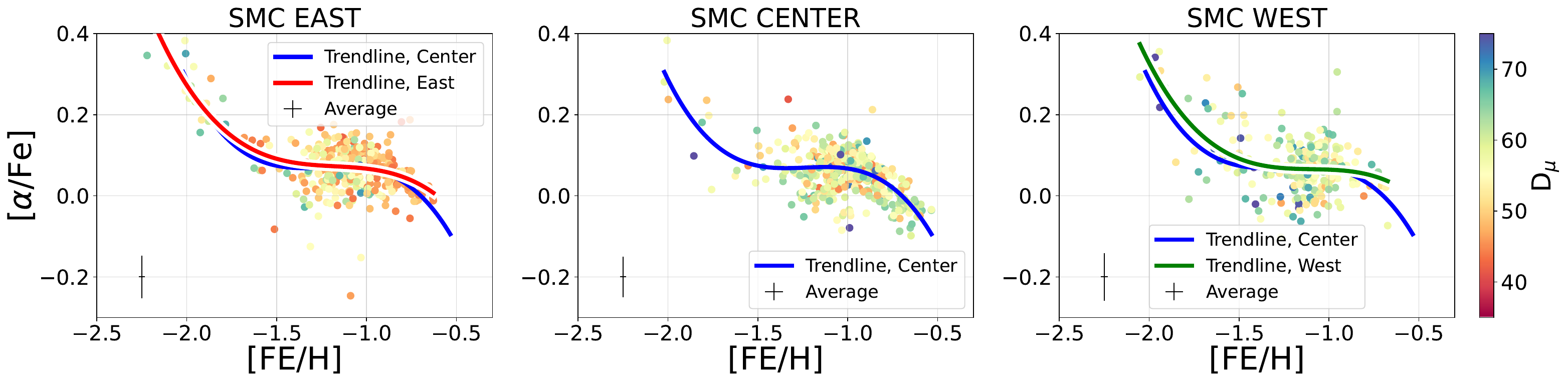}
    \caption{\label{AlphaFe3} 
    The distribution of [$\alpha$/Fe]-[Fe/H] 
    for the three primary SMC regions, with points color-coded by stellar distance. It is evident that the West region lacks the metal-rich, low-[$\alpha$/Fe] population evident in the East and Center regions.
}
\end{figure*}


Figure \ref{AllPM-MS} shows
the detailed proper motion vector point diagram
in an SMC-centric coordinate system, and with the points color-coded by stellar NMSU distances. Dramatic differences in the 2-D stellar motions are obvious in this representation.
Two major concentrations of stars having significantly different distances and proper motions are seen in the Eastern region, with the 
stars in the lower right ``lobe'' of the distribution
mostly at a distance of about 40--50 kpc and the stars in the upper left proper motion lobe
mainly at 55--70 kpc.
A similar phenomenon can be seen in the Center region, but with fewer of the closer stars in the lower right ``lobe'' represented.
On the other hand, this bimodal population distribution is virtually absent
in the West region, which shows primarily a more uniform distribution of higher distance stars in the upper left proper motion lobe.

The origin of the bimodal population distributions of distance and proper motion seen in the Eastern and Center regions may reflect the greater proximity of these regions to the LMC and the presence on this side of SMC stars strongly affected by interaction of the SMC and LMC (see below).


\begin{figure}
    \centering
    \includegraphics[scale=0.35]{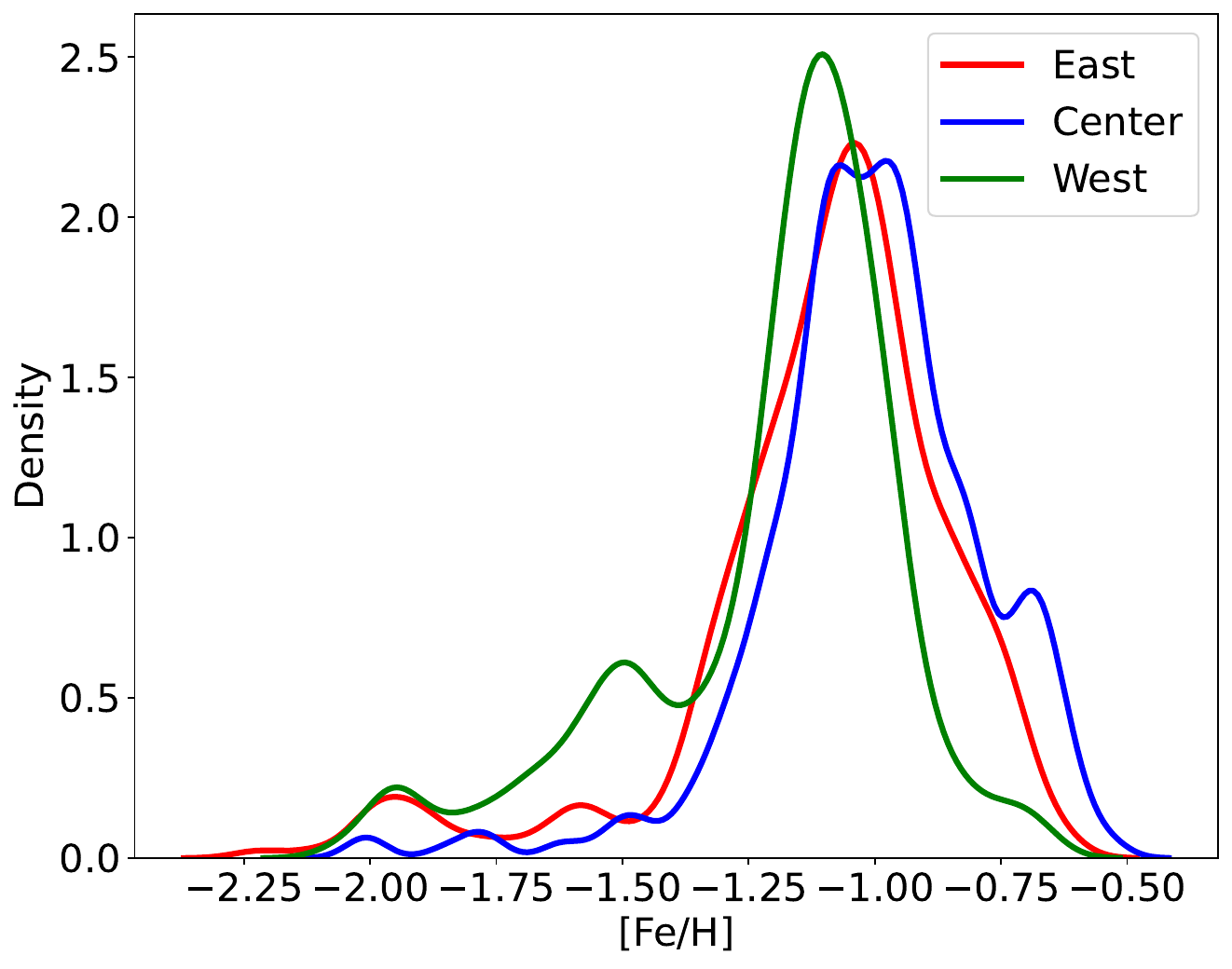}
    \caption{\label{Metalhist3} Metallicity distribution functions, made using a kernel density estimator, for the three primary SMC regions. The Western region seems to be more metal-poor than the other ones by 0.1 dex.}
\end{figure}

\begin{figure*}
    \centering    \includegraphics[scale=0.43]{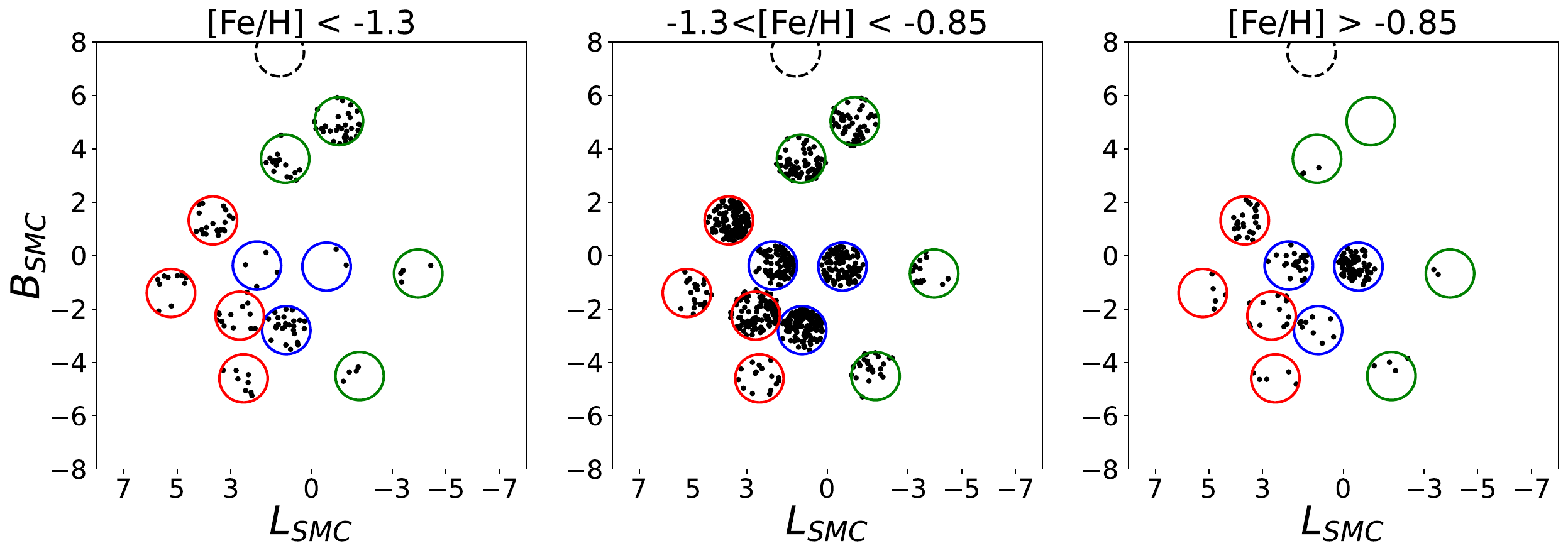}
    \caption{\label{Mapbinned3} Maps of the APOGEE targets, broken into three metallicity groups (``low-'', ``mid-'', and ``high-metallicity''). The West fields show a greater number of low- and mid-metallicity stars, but few stars of high metallicity. The mid- and high-metallicity stars are more centrally concentrated.
    }
\end{figure*}

\subsection{Metallicity and $\alpha$-Abundances}
\label{subsec:metallicity}

Having clearly established that the APOGEE SMC targets show clear proper motion
variation from the galaxy's eastern to its western sides --- likely reflecting the presence of a bimodal population distribution in distance on the East side --- it is important to establish whether the same is observed in the chemical properties of these stars.

Figure \ref{AlphaFe3} shows the APOGEE ASPCAP
[$\alpha$/Fe]-[Fe/H]
abundance distribution of the SMC RGB stars for each of the three primary regions of study.  The overall pattern seen in the three panels is similar 
to that previously shown in \citet{2020ApJ...895...88N}, where there is an overall increase in [$\alpha$/Fe]
as [Fe/H] decreases, all of the way to [Fe/H]$\lesssim-2$ dex. 
The metal-rich end of the overall SMC [Fe/H] distribution reaches [Fe/H]$\sim-0.5$ dex, as also found by Nidever et al.\footnote{While the abundances presented in \citet{2020ApJ...895...88N} were already reliable, the measurements of 
[Fe/H] and [$\alpha$/Fe] presented here are expected to be even more reliable, since they are based on more accumulated $S/N$ in the APOGEE spectra.}


In more detail, however, it can be seen that the general trend shown in the Figure \ref{AlphaFe3} panels is somewhat complex, featuring an inflection region where the overall trend levels out to [$\alpha$/Fe] $\sim +0.05$ in the range $-1.5 \lesssim \rm{[Fe/H]} \lesssim -1.0$. 
Although it can be inferred, this feature is not as clear in the Center region due to the fewer metal-poor stars represented in this overall more metal-rich part of the SMC.
The West region shows more scatter in the $\alpha$-element abundances but exhibits a metal-poor tail similar to the East region, which suggests that there are smaller selection bias differences between the East and West regions.

If this is the case, then the most relevant and striking part of the distribution showing variation between the three panels (and particularly between the East and West regions) in Figure \ref{AlphaFe3} is at the metal-rich end of the distributions.  Here, reminiscent to what is seen in the proper motion distributions shown in Figure
\ref{AllPM-MS}, an ``extra'' metal-rich population is observed in the East and Center regions that is virtually absent in the West region. As shown by the color-coding in the figure, the extra population appearing in the East and Center regions is the same 40--50 kpc population inhabiting the lower-right lobe 
seen in Figure \ref{AllPM-MS}.  Figure \ref{AlphaFe3} shows that this extra population is also relatively metal-rich and has subsolar [$\alpha$/Fe].  Indeed, it is the presence of this extra metal-rich population that is responsible for the downward turn of the inflection seen in the overall trend for [Fe/H] $\gtrsim -1.0$ in the East and Center regions.


\begin{figure*}
    \centering
    \includegraphics[scale=0.3]{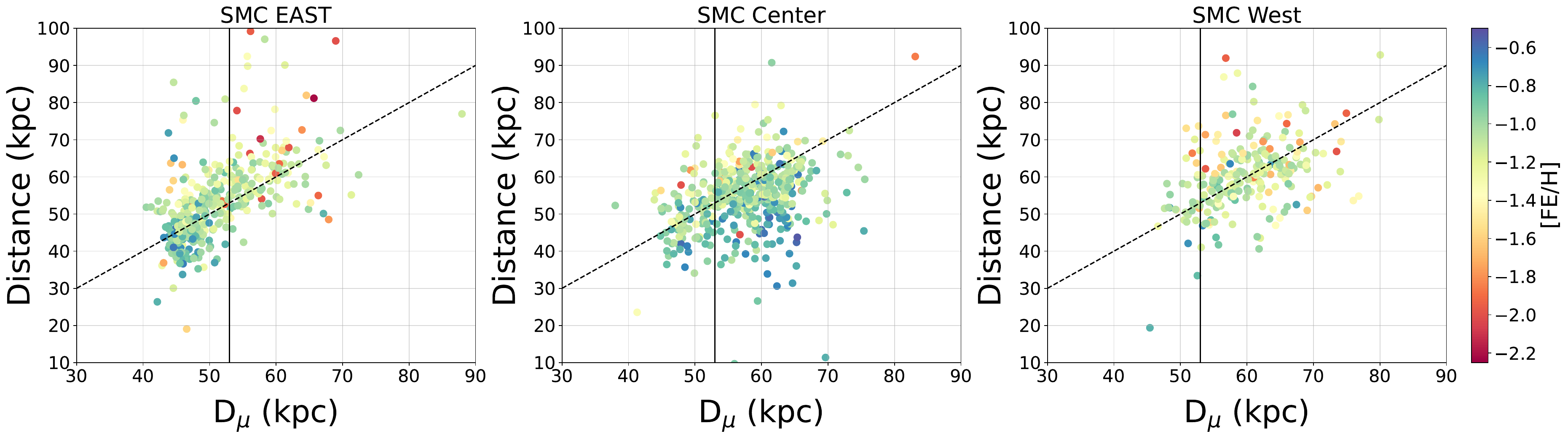}
    \caption{\label{AllTPMdist} The joint distribution of proper motion distance, metallicity, and distance 
    for the three primary regions. The East and Center regions show the presence of a large
    concentration of closer, more metal-rich, low \pmd that are not 
    not present in the West. The dividing line at \pmd = 52 kpc is used for further analysis of the East region in \S\ref{sec:CloserLook}.}
\end{figure*}

These differences in the chemical properties seen across the APOGEE SMC fields are further illustrated in Figures \ref{Metalhist3} and \ref{Mapbinned3}.
Figure \ref{Metalhist3} shows the generic metallicity distribution functions (MDF) for the three primary SMC regions.  Similar MDFs are shown by the East and Center regions, with both showing the same peak at $\sim - 1.0$ dex,
although the Center region is more asymmetrically skewed to the metal-rich end while the East region is more asymmetrically skewed to the metal-poor end, at least in terms of showing a bit stronger representation of [Fe/H] $\lesssim -1.3$ stars.  The West region also shows a peak at [Fe/H] $\sim - 1.0$ dex, but with a distinctly narrower distribution around that peak compared to those seen in the East and Center regions and with a greater fraction of metal-poor stars represented; these two features makes the overall MDF and mean metallicity of the West region lower overall, and to have the appearance of having two populations --- one with a tight metallicity distribution centered on [Fe/H]$ = -1.1$ and a second with a broad metallicity dispersion centered at [Fe/H]$ \sim -1.5$.  Nevertheless, a key feature of the West fields is that they lack the metal-rich component seen clearly in the East and Center fields.  This is made even more clear in Figure \ref{distVpos}, discussed below.

The overall shift to lower mean metallicity in the West compared to East and Center fields is also reflected in the spatial distribution of individual stars in different metallicity bins, as shown in Figure \ref{Mapbinned3}.  This figure also gives the distinct impression that the metal-rich APOGEE SMC targets are 
more tightly concentrated to the SMC center, whereas more metal-poor stars are more broadly dispersed across the SMC, though tending to have a higher representation in peripheral fields. However, it also shows that the most metal-rich stars by and large cover the Center and East regions, but are not represented in the West region.  The observed overall metallicity distribution of the APOGEE SMC stars is made more clear by Figure \ref{distVpos} and comports to the metallicity distributions and gradients previously reported by \citet{Carrera2008}.

\subsection{Joint Distribution of Kinematics, Metallicity, and Distance}
\label{subsec:jointdistribution}

In the preceding sections we have explored more or less separately the distances, kinematics, and chemistry of the APOGEE SMC stars.  Here we look at these properties more holistically.
Figure \ref{AllTPMdist} summarizes the joint distribution of proper motion distance, NMSU distance, and metallicity across the three primary SMC regions of study, with the points color-coded by the latter property.  The observed linear trend of proper motion distance with distance is what one expects for a system of stars sharing a common bulk motion.  However, the extent of that trend is clearly varying across the three primary regions.
As shown previously in Figure
\ref{AllPM-MS},
\pmd spans a wide range in the East region, 
from $\sim$40 kpc to $\sim$70 kpc, while the Center region has \pmd as low
as $\sim$44 kpc.
Stars with \pmd $<$ 52 kpc constitute the majority (61\%) in the East region, but only about a quarter (22\%) of those in the Center.
In contrast, the West has very few stars with \pmd values below 52 kpc, and almost no stars with a distance $<45$ kpc.
Figure \ref{AllVhelioDist} also demonstrates (again, see \S\ref{subsec:metallicity} and Figs.\ \ref{AlphaFe3} and \ref{Metalhist3}) that
the \pmd $<$ 52 kpc 
population in the East region is chemically distinct (i.e., of higher metallicity) from any APOGEE stars observed in the West region. The only other place where such higher metallicity stars are nominally found is in the core of the SMC, as is evident by the concentration of such stars in the center panel of Figure \ref{AllVhelioDist} at the nominal SMC distance ($\sim$60 kpc).  This chemical imprint demonstrates that the difference between the East and West distance trends is {\it not} simply a reflection of the orientation of a non-spherical SMC to the line-of-sight.  
Rather, it strongly suggests that a population of metal-rich stars has been drawn out of the central SMC roughly in the direction of the Sun on the eastern side of the SMC, most likely due to disruption by the LMC-SMC interaction. 

\begin{figure*}
    \centering
    \includegraphics[scale=0.38]{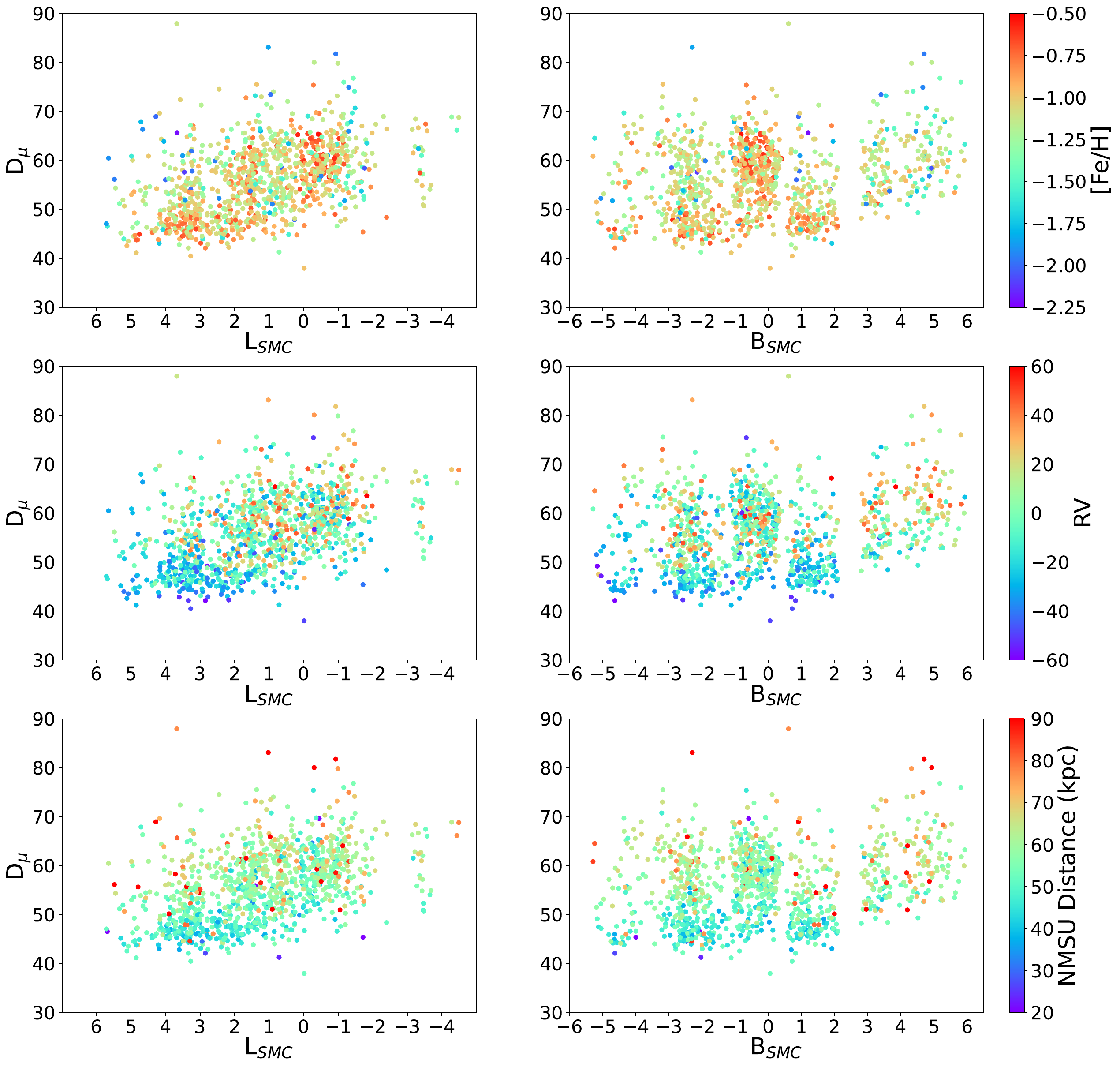}
    \caption{\label{distVpos} Distribution of $D_{\mu}$ as a function of SMC-centric coordinates.  From top to bottom row the points are color-coded by [Fe/H], RV, and NMSU distance, successively.  The Eastern side of the SMC clearly shows an abundant population of approaching, more metal-rich stars on its near side.  The bottom row of panels shows that the NMSU and $D_{\mu}$ distances are correlated. 
 }
\end{figure*}

This suggestion is further illustrated by Figure \ref{distVpos}, which shows the inferred NMSU distances of the APOGEE SMC stars shown as a function of SMC-centric coordinates ($L_{\rm SMC}$, $B_{\rm SMC}$), and with the points color-coded as a function of the various kinematical and chemical properties in hand.  In this coordinate system, the three primary groups of fields we defined (East, Center, West) more or less sort into a sequence with minimal overlap\footnote{The very central SMC3 field does still overlap the West fields SMC1 and SMC10 in this coordinate system.} and in a progression of fields closest to farthest from the Large Magellanic Cloud (left to right).
This figure makes evident the asymmetry of properties across the SMC and that these asymmetries are strongly correlated with the distance of a star from the LMC. In particular, the eastern side of the SMC shows a larger dispersion in all properties surveyed here compared to the western side of the SMC.  This larger dispersion of properties on the eastern side echoes what is also seen in the proper motions \citep[e.g., see Figure 7 of][]{Zivick2018}.
The presence of closer stars on the eastern side of the SMC is evident, as are the trends shown earlier that these stars are more metal-rich and have lower \pmde.  In contrast, these variations in properties with position are not so obvious when plotted in the orthogonal direction (right panels of Fig.\ \ref{distVpos}).
Figure \ref{distVpos} further illustrates how the east-west asymmetry appears to begin with and include the central SMC.  Because the core of the SMC is where the most metal-rich SMC stars are concentrated and likely to have formed, the ``foreground'' population on the eastern side of the SMC may well have originated by the dispersal of SMC core stars towards the Sun.
We further investigate this hypothesis in the next section.





\begin{figure}[h]
    \centering
    \includegraphics[scale=0.38]{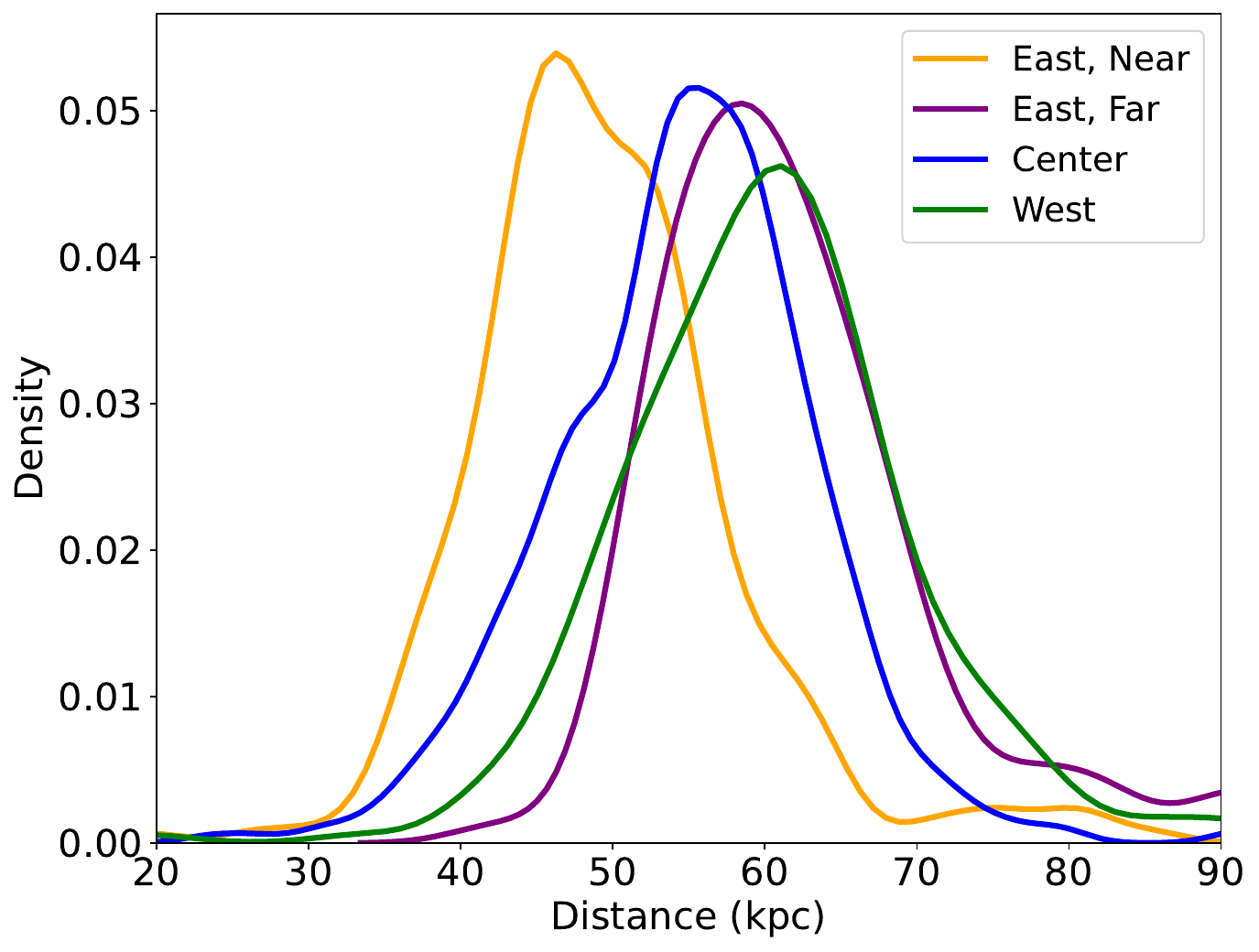}
    \caption{\label{distEast} 
    NMSU distance distribution in the East region, after separation of the two groups using total proper motion criteria (i.e., $D_{\mu}$ --- see Fig.~\ref{AllPMD}). It is clear that the two groups do not share the same peak distance, which supports the notion that there are  
    two populations of RGB stars at different mean proper motion and estimated distance. }
\end{figure}

\section{A Closer Look at the East Region}
\label{sec:CloserLook}

In the previous section, we observed how the APOGEE SMC sample shows a clear asymmetry between its Eastern and Western sides, with the East regions clearly showing evidence of a spread of stars extending to lower \pmde, lower radial velocity, and higher metallicity. In this section, we explore the East region in more detail, to isolate more carefully the properties of the apparent ``foreground'' population appearing there.

\subsection{Distance Division of Populations in the East Region}
\label{subsec:eastdistances}

Based on the analysis of the various properties throughout \S\ref{sec:starproperties}, it seems evident that \pmd is the variable where there is the clearest distinction between the East and West regions (e.g., compare Fig.~\ref{dist-hist3} to
Fig.~\ref{AllPMD}).
We speculate that \pmd provides a better means for isolating the foreground population than the inferred NMSU distance itself because of the greater uncertainties in the latter compared to the uncertainties in the proper motions, which  turn out to be a better proxy for the relative distances of stars sharing a similar bulk motion. 
We also find that a division at \pmd = 52 kpc provides a good criterion for dividing the East region, and isolating the ``foreground'' population (Figs.~\ref{AllPMD}); therefore, we  
adopt this division of the East region stars for the analyses presented throughout Section \ref{sec:CloserLook}.

As a consistency check on this methodology to subdivide the East populations, Figure \ref{distEast} shows the NMSU distances of the East subpopulations divided by the \pmd = 52 kpc criterion. 
It is evident that, as expected, the sub-population with lower \pmd values (the ``Near East'' stars) are systematically closer than the one with higher \pmd (the ``Far East'' stars).
In fact, the Far East sub-population shows a similar NMSU distance distribution to the Center and West regions, which are more clustered around the SMC systemic distance of roughly 60 kpc.

\begin{figure*}
    \centering
    \includegraphics[scale=0.26]{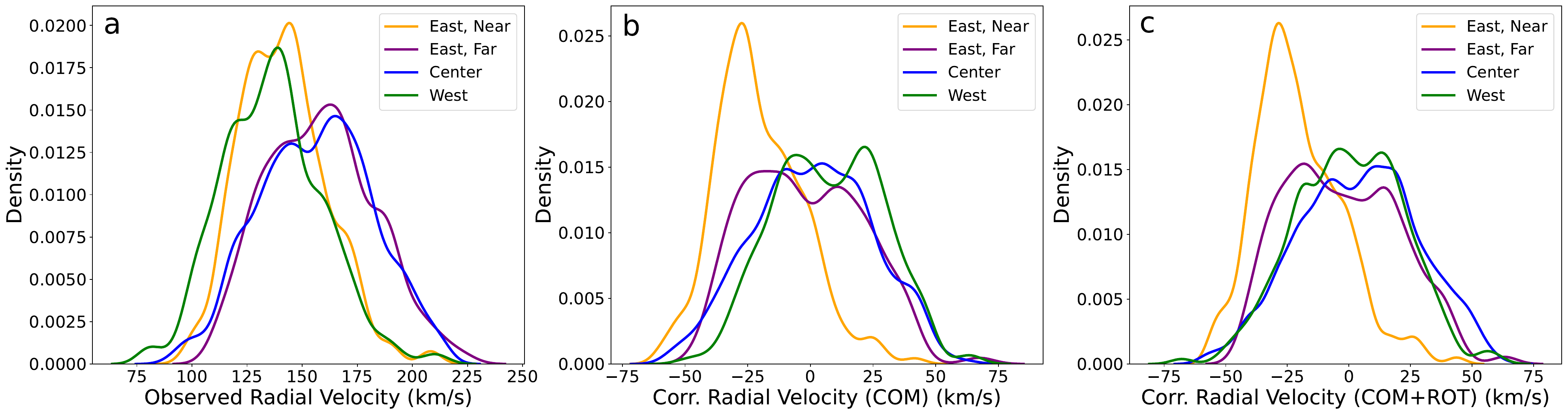}
    \caption{\label{vhelioEast} Radial velocity distributions of the SMC sample, with the East region separated in two near and far subgroups. Panel ({\em a}) shows the raw velocities of the regions as measured by APOGEE. Panel ({\em c}) shows the radial velocity distribution after the correction for bulk SMC motion and rotation described in Section \ref{subsec:rvs}; this correction puts the RVs in the reference frame of the SMC bulk motion. A clear difference in the RV distributions between the Near and Far East sub-populations is evident. The Far East shares an RV distribution more similar to that of the Center and the West regions, while the East Near subpopulation shows a very prominent peak of RVs at around $-$30 \kms not strongly evident in the other regions. 
  }
\end{figure*}

\subsection{Radial Velocity Distributions in the East Region}
\label{subsec:eastrvs}

Figure \ref{vhelioEast} shows the radial velocity distributions of the two East sub-populations along with those of the Center and West regions; panel (a) shows these RV distributions as measured by APOGEE, while panel (b) shows the distributions accounting for the model of bulk SMC motion and rotation described in Section \ref{subsec:rvs}.
Figure \ref{vhelioEast}a turns out to be difficult to interpret for the same reasons it was difficult to interpret Figure \ref{AllVhelio}a. 
The uncorrected RVs for the Near East group show a similar distribution to those of the West region while the Far East Far group RVs closely resemble the Center region RVs. 

On the other hand, correction of the RV distributions for SMC rotation and center of mass movement brings striking clarity to the situation (Figure \ref{vhelioEast}b,c), 
with the Near East sub-population showing a very clear RV peak around $-$30 km s$^{-1}$  and few stars with positive RVs --- an overall distribution that suggests a general flow of stars away from the SMC system.
Meanwhile, the RVs of the Far East group shows a more balanced RV distribution, with a mean of $-2$ km s$^{-1}$, and a general character much more similar to those of the Center and West regions (which themselves, after correction for SMC rotation and systemic motion, show RV distributions that are quite similar to one another).  The shared RV distribution of the Center, West, and Far East populations might be 
interpreted as the general RV distribution of the main body of the SMC.

The distinctly different RV character of the two East subpopulations, and that Far East population shows ``normal'' RVs while the Near East RV distribution is so asymmetrically skewed to motions away from the SMC (and towards us), 
are initial clues as to the origin of the SMC East distance bimodality.
To gain further insights into the origin of these foreground, Near East stars, we next look at the metallicities and abundances provided by APOGEE.


\begin{figure}
    \centering
    \includegraphics[scale=0.38]{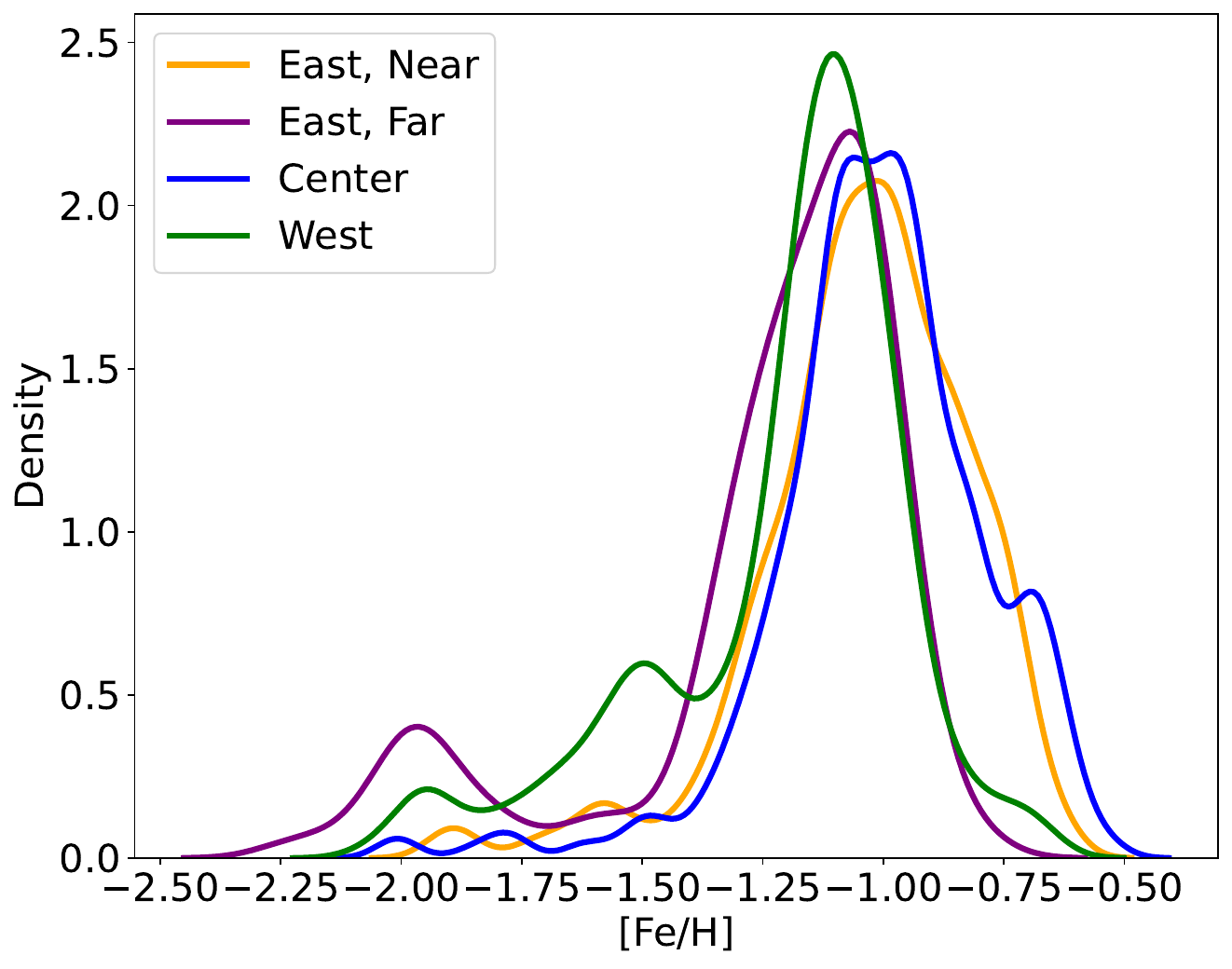}
    \caption{\label{feDist} Metallicity Distribution Functions of the two East region sub-populations along with those of the Center and Western regions. While the distributions for the four populations resemble each other generally, slight differences can be seen in the peak values as well as in the strengths of the low and high metallicity tails.
    In particular, the Near East subpopulation shows an MDF skewed towards higher metallicities just like that for the Center region, whereas the Far East subpopulation shows an MDF with more metal-poor stars and that more closely resembles the MDF of the West region. }
\end{figure}

\subsection{Metallicity and $\alpha$-Abundances in the East Region}
\label{subsec:eastchemistry}


The MDFs for the two East sub-populations provide additional insights into their origins, albeit somewhat more subtly so.
Figure \ref{feDist} compares these MDFs to those of the Center and West regions.  While the four MDFs broadly resemble one another, closer inspection reveals some differences.  First, the Near East MDF is shifted to higher overall metallicities than the Far East MDF, as evidenced by the location of the MDF peaks, but even more so by the asymmetries of the wings of the distributions, with the Far East sub-population showing a skew towards more metal-poor stars and the Near East sub-population showing a skew towards more metal-rich stars.  Moreover, the peak metallicity and metal-poor skew of the Far East sub-population resembles that of the West region, while the MDF --- including the peak metallicity and metal-rich skew --- of the Near East sub-population very closely resembles that of the Center region.  
It is known that, on average, the periphery of the SMC is more metal-poor than the center of the dwarf galaxy \citep[][, Povick et al., in prep.]{Dobbie2014,munoz2023}, so it is perhaps not surprising to see the West and Far East groups reflecting this; but that the Near East population seems to have an MDF very much like that of the Center region is a key feature revealed by our data with potentially strong implications for the origin of the Near East stars.


\begin{figure*}
    \centering
    \includegraphics[scale=0.35]{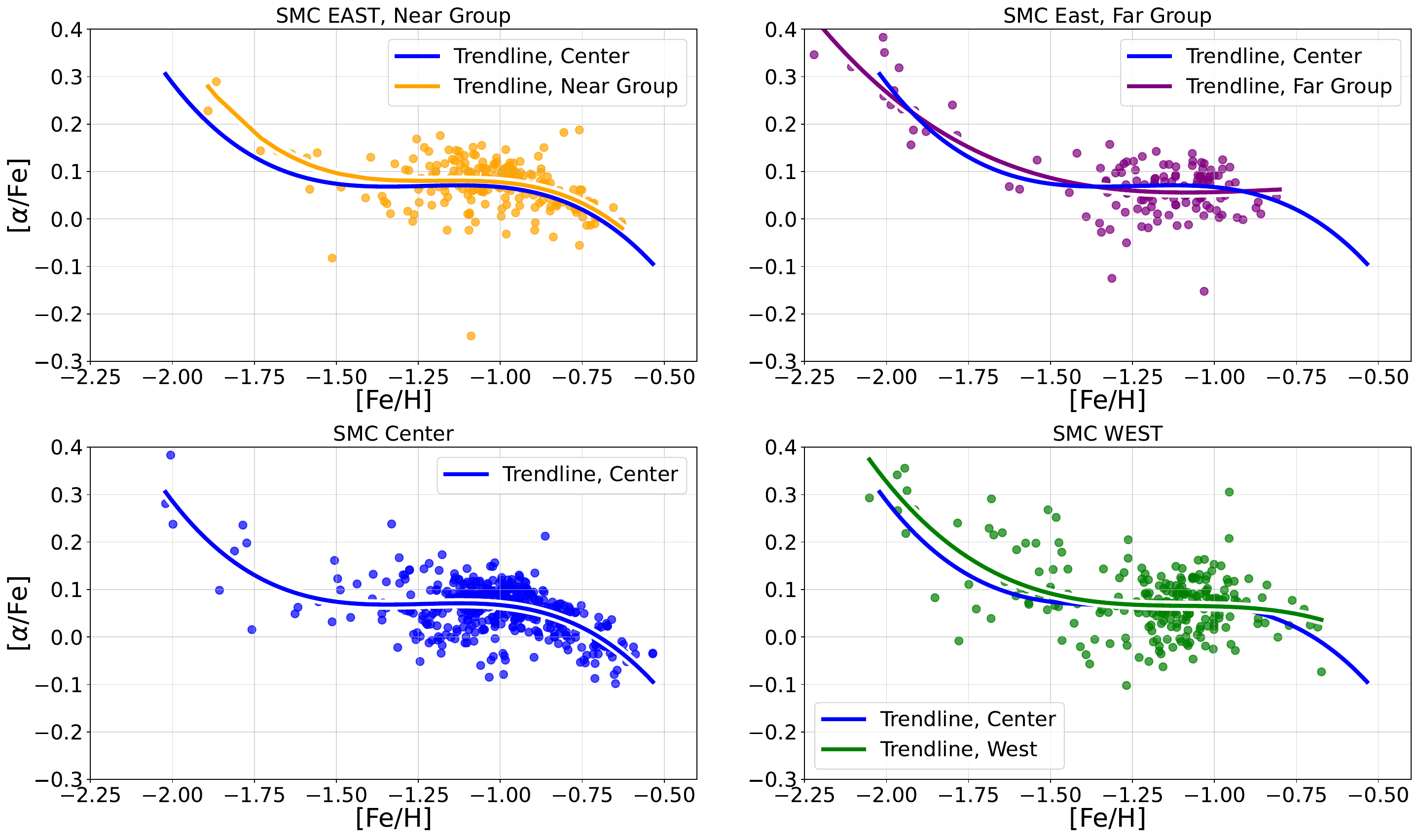}
    \caption{\label{MetalEast} The  [$\alpha$/Fe]-[Fe/H] distributions for the four populations of interest; the two East sub-populations ({\it upper panels}), the Center region ({\it lower left panel}), and the West region ({\it lower right panel}). For each population we show a trendline determined via a bicubic spline.  The horizontal extent of each trendline is determined by the lowest and highest metallicity star in each population.  To highlight the subtle differences between populations --- most especially in the metallicity ranges and MDF --- the trendline for the Center region is included in all four panels.  }
\end{figure*}

Further insights are gained by looking at the detailed chemical abundance pattern distributions. 
Figure \ref{MetalEast} shows the [$\alpha$/Fe]-[Fe/H] distributions of the two East sub-populations along with those of the Center and West regions. We fit a bicubic spline function to each group separately to determine its chemical abundance trendline.  These are shown as purple (Far East), orange (Near East), blue (Center), and green (West) lines in the four panels of Figure \ref{MetalEast}.  The Center trendline (blue) is shown on all panels to facilitate comparison of the distributions for the different SMC populations.  

Broadly speaking, the [$\alpha$/Fe]-[Fe/H] trendlines for the different populations are remarkably similar, suggesting that all populations have participated in the same overall chemical enrichment history.  However, as already shown by the MDFs in Figure \ref{feDist}, the different populations seem to reflect different parts of that enrichment history.  This is made further evident by the [Fe/H] limits of each trendline shown in Figure \ref{MetalEast}, which correspond to the lowest and highest metallicity star in each population, as well as the numbers of stars falling along different parts of each trendline.

In more detail, the Near East sub-population (top left panel of Fig. \ref{MetalEast}) exhibits an $\alpha$-abundance distribution quite similar to the Center region (bottom left panel).  The two trendlines are very similar, with the only discernible variation occuring at the metal-poor end, where the trendline fits are poorly constrained by the low numbers of stars.  While the Center population is slightly more metal-rich than the Near East population and shows a more prominent turnover, both populations are heavily skewed to higher metallicities with relatively few metal-poor stars.\footnote{The fact that the Center population appears to show more metal-poor stars in Figure \ref{MetalEast} is because there are more overall stars sampled in the Center region.  This visual impression is of course diminished in the normalized histograms shown in Figure \ref{feDist}.}
 
In contrast to what is seen in the left panels of Figure \ref{MetalEast}, the [$\alpha$/Fe]-[Fe/H] distribution of the Far East sub-population (top right panel) much more resembles that of the West region (right bottom panel).  While there is more scatter in the [$\alpha$/Fe] values for the stars in the West region, nevertheless, the West and Far East population trendlines are very similar in shape and [Fe/H] extent, and the overall [Fe/H] spreads in the stars of these two populations are very similar (as already shown in Figure \ref{feDist}).
The most striking difference between the Center/Near East versus the West/Far East [$\alpha$/Fe]-[Fe/H] distributions is the lack of metal-rich stars (e.g., [Fe/H]$>$ $-$1.0) 
in the latter populations (as already shown particularly well in Figure \ref{feDist}).  

Overall, it is natural to find more metal-rich 
stars in the central SMC because, as is the case in most galaxies, star formation and chemical evolution proceeded further there.  The lack of such populations in the outskirts of the SMC is also not surprising since these regions do {\em not} currently have much ongoing star formation or gas.  The unexpected result is that the Near East sub-population should exhibit such metal-rich and younger populations.  These are out of place in the periphery, and given their striking chemical resemblance to the stars in the Center region, it seems natural to presume these stars originated more centrally in the SMC. 

\begin{figure*}
    \centering
    \includegraphics[scale=0.4]{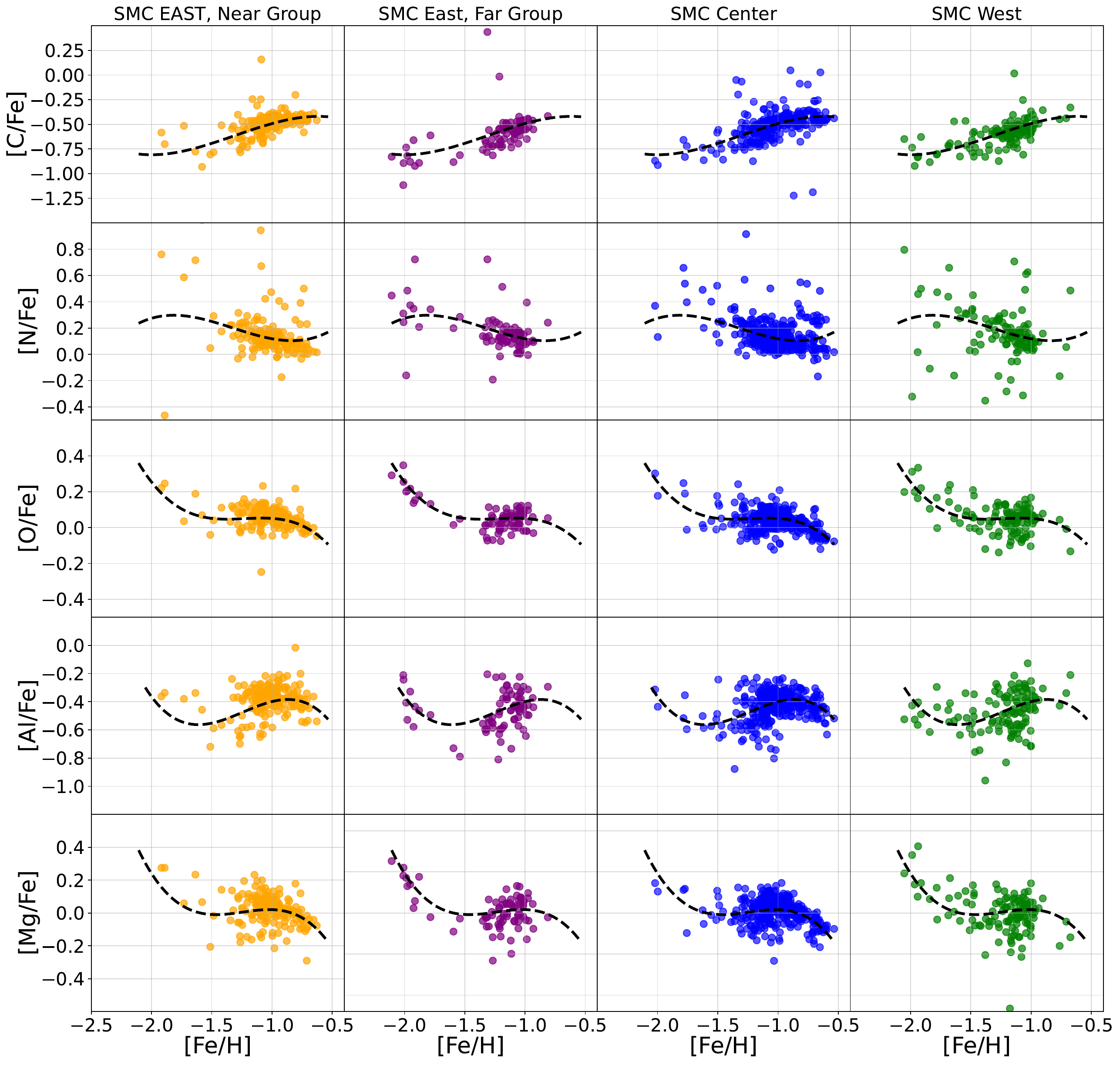}
    \caption{\label{AbundancesCN} Chemical abundance distributions of [Mg/Fe], [Al/Fe], [O/Fe], [N/Fe], and [C/Fe] for the four regions of interest. }
\end{figure*}
\begin{figure*}
    \centering
    \includegraphics[scale=0.4]{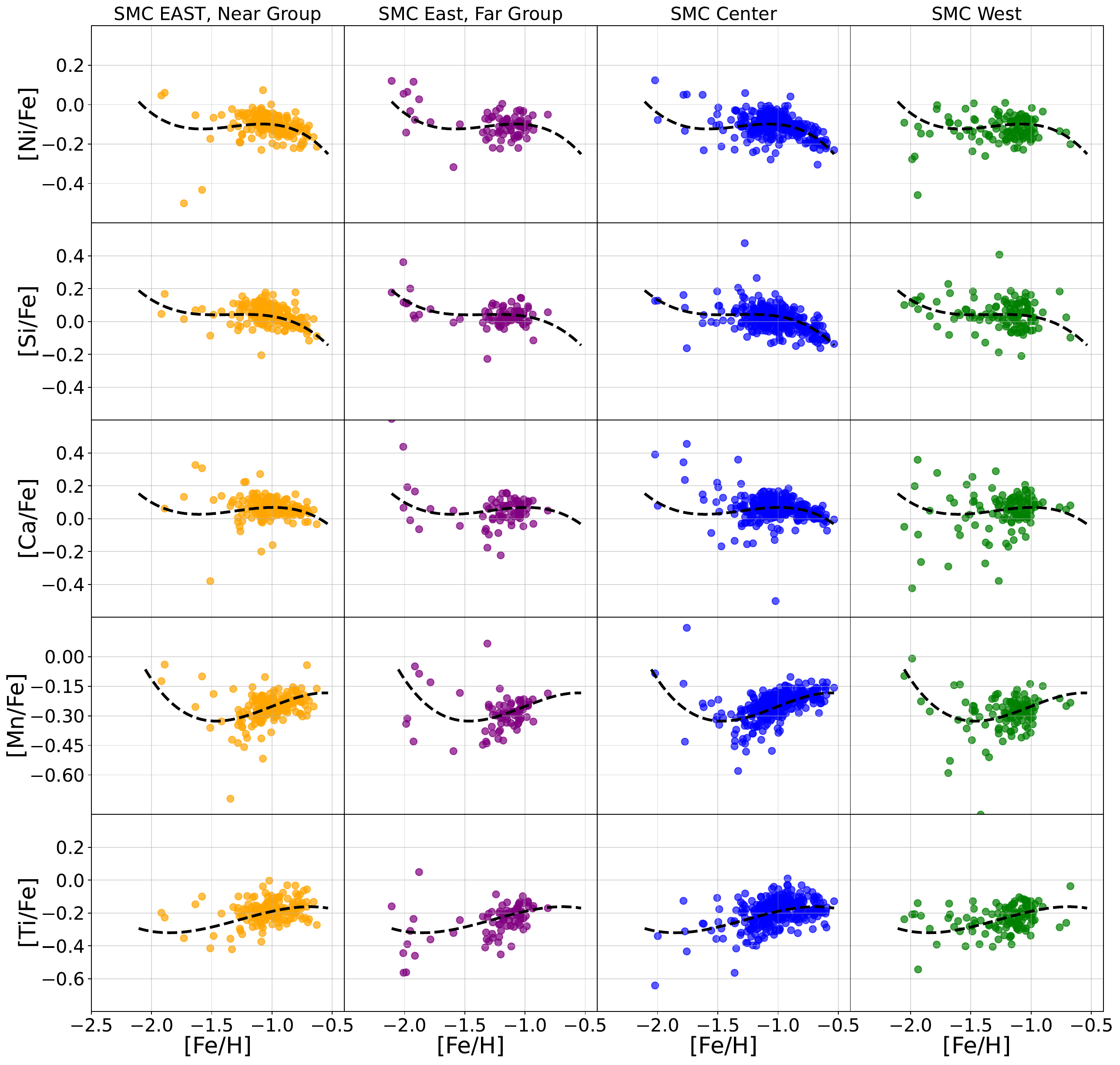}
    \caption{\label{AbundancesOAL} Same as Figure \ref{AbundancesCN} but showing the chemical abundance distributions of [Ti/Fe], [Mn/Fe], [Ca/Fe], [Si/Fe], and [Ni/Fe] for the four regions of interest. }
\end{figure*}

Interestingly, the above observations pertaining to similarities and differences between the [$\alpha$/Fe]-[Fe/H] distributions of the four populations are only repeated, and therefore reinforced, by comparisons in other chemical spaces.
Figures \ref{AbundancesCN} and \ref{AbundancesOAL} 
show similar projections in many other projections of APOGEE's multi-elemental abundance space in an effort to discern any potential signatures or even hints  of differences in the chemical evolution of the two sub-populations of the East region of the SMC. 
As with the [$\alpha$/Fe]-[Fe/H] comparisons, the most striking differences between the populations in all of these chemical spaces comes in the MDF variations already noted.
More significantly, there are no obvious differences in the trendlines of these different chemical spaces across the four SMC populations of the SMC.  
With the numerous chemical species brought to bear, the lack of any trendline differences only strengthens the conclusion that both East sub-populations share the same enrichment histories between them and with the stars of the Center and West regions. 
If there are any differences in the chemical enrichment of the different regions caused by the tidal interaction between the SMC and the LMC, it is still in its early phases and cannot yet be detected.  This is not surprising as the latest close interaction between these dwarf galaxies happened only $\sim$200 Myr ago \citep{Besla2012,Zivick2018}.


\begin{table*}
    \label{tab:table_2D-KS}
    \caption{P-values for 2-dimensional K-S test results. }
\begin{tabular}{cccccc}
    \hline
    \hline
    Element & East Close & East Close & East Close & East Far & East Far\\
     & vs. & vs. & vs. & vs. & vs.  \\
     & East Far & Center & West & Center & West \\
    \hline 
    [$\alpha$/Fe] & 1.00$\times$10$^{-4}$ &  4.01$\times$10$^{-2}$ & 6.24$\times$10$^{-6}$ & 2.47$\times$10$^{-5}$ & 2.20$\times$10$^{-1}$\\
    C/Fe & 5.26$\times$10$^{-9}$ &  1.56$\times$10$^{-1}$ & 7.41$\times$10$^{-13}$ & 3.09$\times$10$^{-8}$ & 3.31$\times$10$^{-1}$    \\
    N & 3.72$\times$10$^{-4}$ &  2.59$\times$10$^{-1}$ & 2.02$\times$10$^{-5}$ & 3.57$\times$10$^{-6}$ & 4.59$\times$10$^{-1}$    \\
    O & 5.01$\times$10$^{-6}$ & 7.06$\times$10$^{-5}$ & 9.74$\times$10$^{-7}$ & 6.90$\times$10$^{-6}$ &  3.11$\times$10$^{-1}$  \\
    Al & 3.51$\times$10$^{-5}$ & 1.82$\times$10$^{-2}$ & 3.40$\times$10$^{-9}$ & 1.64$\times$10$^{-5}$ & 2.61$\times$10$^{-1}$  \\
    Mg & 2.05$\times$10$^{-4}$ & 2.73$\times$10$^{-1}$ & 1.94$\times$10$^{-6}$  & 3.5$\times$10$^{-5}$ & 7.42$\times$10$^{-2}$  \\
    Ni & 4.20$\times$10$^{-4}$ &  1.13$\times$10$^{-1}$ & 3.34$\times$10$^{-6}$ & 1.66$\times$10$^{-5}$ & 1.17$\times$10$^{-1}$   \\
    Si & 8.38$\times$10$^{-5}$ & 2.56$\times$10$^{-6}$ & 4.70$\times$10$^{-6}$ & 1.62$\times$10$^{-5}$ &  1.69$\times$10$^{-1}$    \\
    Ca & 3.16$\times$10$^{-5}$ & 1.87$\times$10$^{-1}$ & 4.08$\times$10$^{-6}$ & 2.63$\times$10$^{-5}$ & 1.68$\times$10$^{-1}$   \\
    Mn & 4.06$\times$10$^{-4}$ & 1.36$\times$10$^{-2}$ &  7.89$\times$10$^{-6}$ & 5.24$\times$10$^{-9}$ & 2.28$\times$10$^{-1}$  \\
    Ti & 9.27$\times$10$^{-6}$ & 6.87$\times$10$^{-2}$ & 2.56$\times$10$^{-6}$ & 2.40$\times$10$^{-6}$ & 2.51$\times$10$^{-1}$   \\
    \hline
\end{tabular}

\end{table*}

\section{Discussion}
\label{sec:discussion}


We have analyzed the characteristics of over 1,000 stars selected to be RGB members of the SMC on the basis of APOGEE-2S spectroscopic stellar atmospheric parameters and RVs combined with {\it Gaia} DR3 proper motions and parallaxes.  These stars lie in eleven APOGEE-2S fields distributed across the face of the SMC and extending to as far as 6 deg from the SMC center. 
We have divided these fields into West, Center, and East regions based on broadly shared spatial and dynamical characteristics of the stars in these fields (Appendix 1); the East region covers the area of the SMC where a line-of-sight distance bimodality has been previously reported \citep{Nidever2013}.  A primary goal of our analysis is to ascertain the origin of this distance bimodality.

Because distances for RGB stars are currently not very precise, we supplemented our analyses by use of a ``proper motion distance'' ($D_{\mu}$), obtained by assuming a bulk transverse velocity for the SMC of 398 km s$^{-1}$, a value chosen to yield a mean distance for the SMC center of 60 kpc.  We found that, while consistent with the spectroscopically based NMSU distances, the $D_{\mu}$ provide more coherent relative distances (i.e., a distribution with less scatter).
With the observation of a larger line-of-sight depth there, we have further subdivided the East region into Near and Far sub-populations at $D_{\mu}=52$ kpc.  This choice was motivated by the asymmetric appearance of the East population distribution in Figure \ref{AllPMD}.  Further investigation of the properties of the two subpopulations created by such a division revealed that a $D_{\mu}=52$ kpc division yielded good chemical and kinematical separation.  Nevertheless, it should be kept in mind that our division into Near and Far East populations is not meant to be definitive for any particular star, but is merely a tool for broadly exploring the bulk properties of the foreground and background populations.
Analysis of the four main SMC populations (West, Center, Near East, and Far East) thus defined has revealed the following clues as to the origin of the East region distance bimodality:

\begin{itemize}
    \item While we do not have distances of the precision to observe the distance bimodality with our data, using proper motions as a proxy for relative distance allows us to observe an asymmetric distance spread in the East region that also yields a mean distance for stars there that is about 10 kpc closer than for the mean of the Center and West regions. (Fig.~\ref{AllPMD}).
    \item The radial velocities, either as observed or corrected for the SMC's projected bulk motion and rotation, show a strong imbalance towards stars flowing away from the SMC (and towards us) for the East region compared to the Center and West regions (Fig.~\ref{AllVhelio}).  The Near East population is dominated by these outward flowing, approaching stars (Fig.~\ref{vhelioEast}).
    \item A smaller sampling of these more nearby and approaching stars are also found to a lesser degree in the Center region (Figs.~\ref{AllVhelioDist} and \ref{distVpos}).  
    \item These closer and approaching stars seen in the Near East population and to a lesser degree in the Center region also show significant differences in relative proper motion compared to the Far East, farther Center, and West populations (Fig.~\ref{AllPM-MS}).  In the SMC ($L$,$B$) coordinate system, stars in the Far East, more distant Center, and West populations show proper motions almost entirely with $\mu_{SMC,B} > -1.4$, whereas the closer, approaching stars in the Near East and closer Center regions generally have $\mu_{SMC,B} < -1.4$.
    \item The MDF of the stars in the Near East population closely resembles that of the Center population in terms of peak metallicity and in having more metal-rich stars and fewer metal-poor stars (Fig.\ref{feDist}).  On the other hand, the Far East MDF resembles that of the West region in terms of peak metallicity and in having a more substantial tail to lower metallicities as well as a lack of metal-rich stars.
    \item Despite the above metallicity differences, all of the populations and subpopulations investigated seem to share the same overall chemical enrichment history, as seen in all element ratio spaces investigated (Figs.~\ref{MetalEast}, \ref{AbundancesCN}, and \ref{AbundancesOAL}).
\end{itemize}

The collected observations suggest a scenario whereby the Far East stars represent the nominal SMC periphery
population, just like the West region stars.  
On the other hand, the spatially and dynamically distinct Near East stars share a similar chemistry to the stars in the Center population, and this ``Near East'' population seems to also be represented to a lesser degree in the foreground of the
SMC Center.

We propose a scenario that accounts for all of the above observations wherein the Near East stars represent material flowing out of the SMC center and this extension of stars
currently sits in the foreground of both the Center and the East fields.  This is most evident in Figures \ref{dist-hist3}, \ref{AllPMD}, \ref{AllVhelio}, \ref{AllVhelioDist}, \ref{AllTPMdist}, \ref{distVpos}, \ref{distEast}, and \ref{vhelioEast}.  Such a stellar outflow is a natural product of tidal disruption, of the type expected by the interaction of the SMC with the LMC.
That these foreground stars are extratidal is borne out by the fact that the current tidal radius of the SMC can be estimated to be 3.4--5.1 kpc\footnote{This tidal radius is estimated using an SMC mass of $3-10\times10^9$ M$_{\odot}$ \citep{Gardiner1996,Bekki2009}, a current Milky Way-SMC separation of 57 kpc \citep{Filipovic1996},
and a Milky Way mass to that distance of $\sim 4.6\times10^{12}$ M$_{\odot}$ \citep{Irrgang2013}.}, which is much smaller than the observed spread of stellar distances in the East region and smaller than the separation of the peak estimated distance  for the Near East stars compared to the Far East stars (Fig.~\ref{distEast}).

This proposed scenario builds on a rich heritage of efforts to explain long-recognized structural peculiarities around the SMC, and, particularly, on its Eastern side.  For example, \citet{Shapley1940} pointed out that in deep photographic images taken as far back as 1909 ``a large elliptical extension, or wing, of the Small Cloud of Magellan'' is visible and extending towards the LMC to at least 6.5$^{\circ}$ from the center of the SMC.  Comparing the location of our fields (Figure \ref{fig:fields}) to traditional density maps of this ``SMC Wing'' (see, e.g., Fig. 3 of \citealt{Westerlund1971} or Fig. 5 of \citealt{Albers1987}) shows that several of our East fields (SMC7 and SMC11) overlap this originally identified surface brightness feature.  This stellar extension was recognized as ``very clearly a tidal structure pointed at the LMC'' at least as far back as \citet{Caldwell1992}.  However, it is worth pointing out that this traditional SMC Wing was observed using blue, presumably young stars.


This long-known young stellar structure has since been associated with the HI Magellanic Bridge \citep[e.g.,][]{Hindman1961,Putman1998,Muller2003}, a feature widely considered to be tidally stripped gas from the SMC due to a close encounter with the LMC some 200 Myr ago
\citep[e.g.,][]{Gardiner1996,Muller2007,Diaz2012}. Though stars that have been identified with this tidally stripped gas are predominantly young ($<$200 Myr), intermediate-old aged ($\sim$1-12 Gyr) stellar populations exist there too \citep{Hatzidimitriou1989,Nidever2013}.  The latter have been shown to be extended along the line of site based on reliable standard candles, like red clump stars \citep{Hatzidimitriou1989,Gardiner1991}, which were used to show the clear bimodality on the eastern side of the SMC, with one concentration of stars --- ``the eastern stellar structure'' --- having a mean distance as close as $\sim$55 kpc from the Sun \citep{Nidever2013}. Taken together, the collected data 
are strengthening the picture that a tidal arm of stripped stars and gas extends from the front side of the SMC and towards the LMC, consistent with models producing tidally stripped material from a recent LMC-SMC interaction.

We submit that what we identify here as the ``Near East'' population corresponds to ``the eastern stellar structure'' identified by \citep{Nidever2013}.  The fact that the latter authors put the distance of this population at 55 kpc may at first seem inconsistent with the $D_{\mu}<52$ kpc criterion we adopt to define the Near East population, but this apparent inconsistency simply reflects the fact that our available RGB distances are not as reliable as RC distances and that $D_{\mu}$ was developed as a proxy for {\it relative} distances; indeed, the mean NMSU distances for these RGB stars are more consistent with the Nidever et al. mean RC distance  to ``the eastern stellar structure'' (Fig.~\ref{distVpos}).

Our new analysis adds to the emerging picture of the overall SMC structure by, first, definitively confirming --- via multi-element chemical abundances (Figs.~\ref{MetalEast}, \ref{AbundancesCN}, and \ref{AbundancesOAL}) --- that the foreground population of the \citet{Nidever2013} ``bimodality'' on the east side of the SMC is in fact coming from the SMC.\footnote{By the way, comparison of the upper left panel of Figure \ref{MetalEast} to the [$\alpha$/Fe]-[Fe/H] distribution for the LMC shown in Figure 14 of \citet{Nidever2020} shows that the foreground stars are not from the LMC, which shows a different trendline in this chemical space at the metal-poor end and a strong skew of its MDF to a higher metallicity, reaching some 0.4 dex more metal-rich than seen in the SMC foreground population.  In addition, \citet{Nidever2013} showed that the density of the ``eastern stellar structure'' {\em increases} as the radius to the SMC decreases, which is inconsistent with an LMC origin.} Moreover, while the distances of our RGB stars are not as reliable as those for RC stars, we provide further information on the 3-D configuration of the foreground SMC tidal extension in that we show that it appears to lie in the foreground of the SMC both across its eastern side as well as across much of the center (Fig.~\ref{distVpos}).
Furthermore, the impression given by Fig.~\ref{distVpos} is that this population may actually {\it emerge} from the central SMC and get progressively nearer (i.e., farther from the SMC) as it reaches eastward towards the LMC.  A connection to the central SMC is strengthened by the more metal-rich MDF of the stars in the Near East population, which strongly matches that of the central SMC.  Given that conventional models of tidal disruption generally show that tidal stripping tends to act most strongly on the least bound stars on the periphery of a satellite, it is perhaps counterintuitive that outflows of material might come from the {\it center} of the SMC. However, this may merely attest to the strength and small impact parameter of the recent LMC-SMC interaction.  Indeed, some models suggests this was a dramatic impact, with an impact parameter as small as 5 kpc or less \citep{Besla2012,Besla2013,Zivick2018,Choi2022} --- surely close enough to dramatically disrupt the central part of the SMC.


Our data also provide new and strong kinematical constraints on LMC-SMC interaction models by providing the RVs of the stars in the foreground tidal extension (Fig.~\ref{vhelioEast}). The modal velocity of the stars flowing from 
the SMC appears to be about $-$30 km s$^{-1}$ relative to the SMC center, although higher velocities are also seen (Fig.~\ref{vhelioEast}b).
Taken together with the proper motions, these RV data will allow more refined 6-D phase-space models of the SMC to be developed. It has already been shown \citep{Zivick2019} that the SMC has a complex distribution of proper motions that cannot be explained by simple rotation models.  Indeed, \citet{Zivick2019} needed to include a tidal expansion in their model to account for this complexity.  The strong RV asymmetry observed in the SMC (Fig.~\ref{vhelioEast}) due to ``the eastern stellar structure'' may well account for this tidal expansion term.  More importantly, future SMC models will need to include a foreground tidal arm with substantial stellar mass and density. For example, in the East field, the number of stars in the Near population ($\sim$226) outnumbers those in the Far population ($\sim$139) by a factor of $\sim1.6$; thus the foreground population, at 62\% of the stellar surface density, dominates the East periphery of the SMC.

The presence of a strong tidal feature to one side of the SMC has no doubt influenced attempts to interpret or fit its 3-D structure using conventional galaxy models. 
A variety of standard candle tracers --- e.g., Cepheid, red clump, and RR Lyrae stars --- have shown strong evidence for closer distances to the east that suggest the SMC to be highly elongated with a high inclination angle and/or suggestions of a barred morphology \citep{Caldwell1986,Laney1986,Caldwell1992,Haschke2012, Nidever2013,Jacyszyn-Dobrzeniecka2017}.  How much of that elongation is due to tidal stretching or debris versus the intrinsic, bound shape of the unperturbed progenitor system remains to be resolved.



This paper has explored tantalizing evidence for a largescale gravitational perturbation of the SMC that is manifested as tidal debris pulled out primarily from the center of the SMC and lying in the foreground of our view of the center to the eastern side of the SMC.  This proposition can be tested by a variety of future observations and measurements. For example, it would suggest that the foreground stars on the eastern side of the SMC should be found to be younger on average than the more distant stars, given that the mean age of stars in the center of the SMC are younger than in its periphery \citep{harriszaritsky2004,rubele2018,massana2022}.  Moreover, the APOGEE coverage of the SMC, while broad, is still admittedly spotty (Fig.~\ref{fig:fields}). More contiguous and precise three-dimensional mapping of the SMC should allow the true shape of the putative tidal arm of debris to be discerned. Fortunately, an extension of the APOGEE survey of the Magellanic Clouds will be undertaken in SDSS-V as part of the Magellanic Genesis Survey (MGS),
a project that will sample tens of thousands of bright stars across the face of the SMC using both the APOGEE-South infrared and BOSS optical spectrographs.  In addition, the 4MOST ``One Thousand and One Magellanic Fields'' \citep[1001MC;][]{Cioni2019} survey will collect spectroscopic data on half a million stars across the Magellanic Clouds.
With MGS and 1001MC, it should be possible to 
assess the size, shape, and extent of foreground SMC tidal structure.

\section*{Acknowledgements}

%
A.A. and S.R.M. acknowledge funding from the National Science Foundation (NSF) grant AST-1909497, while D.L.N. and J.T.P. acknowledge support from NSF grant AST-1908331.  A.A. also acknowledges financial support from the SDSS-V collaboration while working on this project. A.M. gratefully acknowledges support by the ANID BASAL project FB210003,  FONDECYT Regular grant 1212046, and funding from the Max Planck Society through a ``PartnerGroup'' grant

Funding for the Sloan Digital Sky 
Survey IV has been provided by the 
Alfred P. Sloan Foundation, the U.S. 
Department of Energy Office of 
Science, and the Participating 
Institutions. 

SDSS-IV acknowledges support and 
resources from the Center for High 
Performance Computing  at the 
University of Utah. The SDSS 
website is www.sdss.org.

SDSS-IV is managed by the 
Astrophysical Research Consortium 
for the Participating Institutions 
of the SDSS Collaboration including 
the Brazilian Participation Group, 
the Carnegie Institution for Science, 
Carnegie Mellon University, Center for 
Astrophysics | Harvard \& 
Smithsonian, the Chilean Participation 
Group, the French Participation Group, 
Instituto de Astrof\'isica de 
Canarias, The Johns Hopkins 
University, Kavli Institute for the 
Physics and Mathematics of the 
Universe (IPMU) / University of 
Tokyo, the Korean Participation Group, 
Lawrence Berkeley National Laboratory, 
Leibniz Institut f\"ur Astrophysik 
Potsdam (AIP),  Max-Planck-Institut 
f\"ur Astronomie (MPIA Heidelberg), 
Max-Planck-Institut f\"ur 
Astrophysik (MPA Garching), 
Max-Planck-Institut f\"ur 
Extraterrestrische Physik (MPE), 
National Astronomical Observatories of 
China, New Mexico State University, 
New York University, University of 
Notre Dame, Observat\'ario 
Nacional / MCTI, The Ohio State 
University, Pennsylvania State 
University, Shanghai 
Astronomical Observatory, United 
Kingdom Participation Group, 
Universidad Nacional Aut\'onoma 
de M\'exico, University of Arizona, 
University of Colorado Boulder, 
University of Oxford, University of 
Portsmouth, University of Utah, 
University of Virginia, University 
of Washington, University of 
Wisconsin, Vanderbilt University, 
and Yale University.

\section*{Data Availability}

The APOGEE data underlying this article are publicly available via SDSS-IV Data Release 17, available at https://www.sdss4.org/dr17/.  
The Gaia DR3 data used in this article are available at https://gea.esac.esa.int/archive/.

\appendix
\section{Sorting APOGEE SMC Stars into Groups}
The goal of this paper is to explore stars in the periphery of the Small Magellanic Cloud, with a focus on understanding sub-populations therein, and, in particular, understanding the relations of those sub-populations with each other and the populations in the main body of the SMC.  To the end of identifying distinct sub-populations, particularly those that represent the region of the bimodality identified in the SMC periphery, we sought to look for 
sensible groupings based on shared spatial and kinematical characteristics.  To that end, 
we looked at the distributions of stellar heliocentric distance, radial velocity ($V_{helio}$), and total proper motion (\tpme) across the entire APOGEE SMC sample, and divided each of these distributions near the median in each property.  The actual values used in each case are 55 kpc, 150 \kmse, and 1.55 milliarcsec year$^{-1}$.  

Then, for each APOGEE field, we counted the number of stars falling below or above the dividing value in each property (distance, $V_{\rm helio}$, \tpme), assigning stars a score of $-1$ or $+1$, respectively, for each property.  Then, the scores for each property are summed together for each field and the result normalized by the number of stars in each field, to define a total score for each field.  
\begin{table*}
    \label{tab:spec_sum}
    \caption{Details of the scoring system used to group fields into main SMC regions.}
\begin{tabular}{cccccc}
    \hline
    \hline
    Field & Number & Distance Index & $V_{\rm helio}$ Index & $\mu$ Index & Score\\
    \hline
    4 & 165 &$-$0.26 & $-$0.55 & $-$0.15 & $-$0.96\\
    1 & 34 &$-$0.29 & $-$0.18 & $-$0.35 & $-$0.82\\
    2 & 89 &$-$0.64 & 0.78 & $-$0.80 & $-$0.66\\
    5 & 118 &$-$0.12 & $-$0.10 & $-$0.29 & $-$0.51\\
    12 & 24 &$-$0.25 & 0.42 & $-$0.67 & $-$0.5\\
    10 & 93 &$-$0.42 & 0.44 & $-$0.42 & $-$0.4\\
    3 & 183 & 0.21 & 0.19 & $-$0.72 & $-$0.32\\
    11 & 129 & 0.24 & $-$0.16 & 0.35 & 0.43\\
    8 & 29 & 0.31 & $-$0.17 & 0.53 & 0.66\\
    7 & 39 & 0.38 & $-$0.03 & 0.49 & 0.84\\
    6 & 171 & 0.17 & 0.46 & 0.43 & 1.06\\
    \hline
\end{tabular}

\end{table*}

         
    
These scores, shown in rank order in Table \ref{tab:spec_sum}1, show very clearly that fields SMC6, 7, 8 and 11 --- the only ones with positive scores --- are distinct.  These ``East fields'' just so happen to be the four lying closest to the LMC.  This scoring system also demonstrates that the next two fields closest to the LMC, fields SMC4 and 5, contain APOGEE targets with very different spatio-kinematical properties than those in the ``East'' fields.  Given their smaller angular separation to the SMC center (Fig.~\ref{fig:fields}), these two fields, along with field SMC3 (which also has a negative total score), naturally form a ``Center'' field group.  The remaining four fields, SMC1, 2, 10, and 12, have similar angular separations from the SMC center as the East group, and therefore represent an SMC periphery control sample; because of their collective location, we refer to these as the ``West'' fields.  The result of this analysis is that we have sorted the stars in our SMC sample into three similarly-sized groups by their shared distance and kinematical properties, but these groups also happen to divide logically on the sky (Fig.~\ref{fig:fields}).



 


\bibliographystyle{aasjournals}
\bibliography{ref_og.bib}

\begin{thebibliography}{}
\expandafter\ifx\csname natexlab\endcsname\relax\def\natexlab#1{#1}\fi
\providecommand{\url}[1]{\href{#1}{#1}}

\bibitem[{{Albers} {et~al.}(1987){Albers}, {MacGillivray}, {Beard}, \&
  {Chromey}}]{Albers1987}
{Albers}, H., {MacGillivray}, H.~T., {Beard}, S.~M., \& {Chromey}, F.~R. 1987,
  \aap, 182, L8

\bibitem[{{Allende Prieto} {et~al.}(2006){Allende Prieto}, Beers, Wilhelm,
  Newberg, Rockosi, Yanny, \& Lee}]{AllendePrieto2006}
{Allende Prieto}, C., Beers, T.~C., Wilhelm, R., {et~al.} 2006, AJ, 636, 804.
\newblock \url{http://adsabs.harvard.edu/abs/2006ApJ...636..804A}

\bibitem[{{Beaton} {et~al.}(2021){Beaton}, {Oelkers}, {Hayes}, {Covey},
  {Chojnowski}, {De Lee}, {Sobeck}, {Majewski}, {Cohen},
  {Fern{\'a}ndez-Trincado}, {Longa-Pe{\~n}a}, {O'Connell}, {Santana},
  {Stringfellow}, {Zasowski}, {Aerts}, {Anguiano}, {Bender}, {Ca{\~n}as},
  {Cunha}, {Donor}, {Fleming}, {Frinchaboy}, {Feuillet}, {Harding},
  {Hasselquist}, {Holtzman}, {Johnson}, {Kollmeier}, {Kounkel}, {Mahadevan},
  {Price-Whelan}, {Rojas-Arriagada}, {Rom{\'a}n-Z{\'u}{\~n}iga}, {Schlafly},
  {Schultheis}, {Shetrone}, {Simon}, {Stassun}, {Stutz}, {Tayar}, {Teske},
  {Tkachenko}, {Troup}, {Albareti}, {Bizyaev}, {Bovy}, {Burgasser}, {Comparat},
  {Downes}, {Geisler}, {Inno}, {Manchado}, {Ness}, {Pinsonneault}, {Prada},
  {Roman-Lopes}, {Simonian}, {Smith}, {Yan}, \& {Zamora}}]{Beaton2021}
{Beaton}, R.~L., {Oelkers}, R.~J., {Hayes}, C.~R., {et~al.} 2021, \aj, 162, 302

\bibitem[{{Bekki} \& {Stanimirovi{\'c}}(2009)}]{Bekki2009}
{Bekki}, K., \& {Stanimirovi{\'c}}, S. 2009, \mnras, 395, 342

\bibitem[{{Belokurov} \& {Erkal}(2019)}]{Belokurov2019}
{Belokurov}, V.~A., \& {Erkal}, D. 2019, \mnras, 482, L9

\bibitem[{{Besla} {et~al.}(2013){Besla}, {Hernquist}, \& {Loeb}}]{Besla2013}
{Besla}, G., {Hernquist}, L., \& {Loeb}, A. 2013, \mnras, 428, 2342

\bibitem[{{Besla} {et~al.}(2007){Besla}, {Kallivayalil}, {Hernquist},
  {Robertson}, {Cox}, {van der Marel}, \& {Alcock}}]{Besla2007}
{Besla}, G., {Kallivayalil}, N., {Hernquist}, L., {et~al.} 2007, \apj, 668, 949

\bibitem[{{Besla} {et~al.}(2012){Besla}, {Kallivayalil}, {Hernquist}, {van der
  Marel}, {Cox}, \& {Kere{\v s}}}]{Besla2012}
---. 2012, \mnras, 421, 2109

\bibitem[{{Blanton} {et~al.}(2017){Blanton}, {Bershady}, {Abolfathi},
  {Albareti}, {Allende Prieto}, {Almeida}, {Alonso-Garc{\'{\i}}a}, {Anders},
  {Anderson}, {Andrews}, \& et~al.}]{Blanton2017}
{Blanton}, M.~R., {Bershady}, M.~A., {Abolfathi}, B., {et~al.} 2017, \aj, 154,
  28

\bibitem[{{Bowen} \& {Vaughan}(1973)}]{bv73}
{Bowen}, I.~S., \& {Vaughan}, A.~H., J. 1973, \ao, 12, 1430

\bibitem[{{Caldwell} \& {Coulson}(1986)}]{Caldwell1986}
{Caldwell}, J.~A.~R., \& {Coulson}, I.~M. 1986, \mnras, 218, 223

\bibitem[{{Caldwell} \& {Meuder}(1992)}]{Caldwell1992}
{Caldwell}, J. A.~R., \& {Meuder}, D.~L. 1992, in Astronomical Society of the
  Pacific Conference Series, Vol.~30, Variable Stars and Galaxies, in honor of
  M. W. Feast on his retirement, ed. B.~{Warner}, 173

\bibitem[{{Carrera} {et~al.}(2008){Carrera}, {Gallart}, {Aparicio}, {Costa},
  {M{\'e}ndez}, \& {No{\"e}l}}]{Carrera2008}
{Carrera}, R., {Gallart}, C., {Aparicio}, A., {et~al.} 2008, \aj, 136, 1039

\bibitem[{{Choi} {et~al.}(2022){Choi}, {Olsen}, {Besla}, {van der Marel},
  {Zivick}, {Kallivayalil}, \& {Nidever}}]{Choi2022}
{Choi}, Y., {Olsen}, K. A.~G., {Besla}, G., {et~al.} 2022, \apj, 927, 153

\bibitem[{{Choi} {et~al.}(2018){Choi}, {Nidever}, {Olsen}, {Blum}, {Besla},
  {Zaritsky}, {van der Marel}, {Bell}, {Gallart}, {Cioni}, {Johnson}, {Vivas},
  {Saha}, {de Boer}, {No{\"e}l}, {Monachesi}, {Massana}, {Conn},
  {Martinez-Delgado}, {Mu{\~n}oz}, \& {Stringfellow}}]{Choi2018a}
{Choi}, Y., {Nidever}, D.~L., {Olsen}, K., {et~al.} 2018, \apj, 866, 90

\bibitem[{{Cioni} {et~al.}(2019){Cioni}, {Storm}, {Bell}, {Lemasle},
  {Niederhofer}, {Bestenlehner}, {El Youssoufi}, {Feltzing},
  {Gonz{\'a}lez-Fern{\'a}ndez}, {Grebel}, {Hobbs}, {Irwin}, {Jablonka}, {Koch},
  {Schnurr}, {Schmidt}, \& {Steinmetz}}]{Cioni2019}
{Cioni}, M. . R.~L., {Storm}, J., {Bell}, C.~P.~M., {et~al.} 2019, The
  Messenger, 175, 54

\bibitem[{{Cioni} {et~al.}(2011){Cioni}, {Clementini}, {Girardi}, {Guandalini},
  {Gullieuszik}, {Miszalski}, {Moretti}, {Ripepi}, {Rubele}, {Bagheri},
  {Bekki}, {Cross}, {de Blok}, {de Grijs}, {Emerson}, {Evans}, {Gibson},
  {Gonzales-Solares}, {Groenewegen}, {Irwin}, {Ivanov}, {Lewis}, {Marconi},
  {Marquette}, {Mastropietro}, {Moore}, {Napiwotzki}, {Naylor}, {Oliveira},
  {Read}, {Sutorius}, {van Loon}, {Wilkinson}, \& {Wood}}]{VMC2011}
{Cioni}, M. R.~L., {Clementini}, G., {Girardi}, L., {et~al.} 2011, \aap, 527,
  A116

\bibitem[{{Cullinane} {et~al.}(2022{\natexlab{a}}){Cullinane}, {Mackey}, {Da
  Costa}, {Erkal}, {Koposov}, \& {Belokurov}}]{Cullinane2022a}
{Cullinane}, L.~R., {Mackey}, A.~D., {Da Costa}, G.~S., {et~al.}
  2022{\natexlab{a}}, \mnras, 510, 445

\bibitem[{{Cullinane} {et~al.}(2022{\natexlab{b}}){Cullinane}, {Mackey}, {Da
  Costa}, {Erkal}, {Koposov}, \& {Belokurov}}]{Cullinane2022b}
---. 2022{\natexlab{b}}, \mnras, 512, 4798

\bibitem[{{Cunha} {et~al.}(2017){Cunha}, {Smith}, {Hasselquist}, {Souto},
  {Shetrone}, {Allende Prieto}, {Bizyaev}, {Frinchaboy},
  {Garc{\'{\i}}a-Hern{\'a}ndez}, {Holtzman}, {Johnson}, {J{\H o}nsson},
  {Majewski}, {M{\'e}sz{\'a}ros}, {Nidever}, {Pinsonneault}, {Schiavon},
  {Sobeck}, {Skrutskie}, {Zamora}, {Zasowski}, \&
  {Fern{\'a}ndez-Trincado}}]{Cunha2017}
{Cunha}, K., {Smith}, V.~V., {Hasselquist}, S., {et~al.} 2017, \apj, 844, 145

\bibitem[{{Dark Energy Survey Collaboration} {et~al.}(2016){Dark Energy Survey
  Collaboration}, {Abbott}, {Abdalla}, {Aleksi{\'c}}, {Allam}, {Amara},
  {Bacon}, {Balbinot}, {Banerji}, {Bechtol}, {Benoit-L{\'e}vy}, {Bernstein},
  {Bertin}, {Blazek}, {Bonnett}, {Bridle}, {Brooks}, {Brunner}, {Buckley-Geer},
  {Burke}, {Caminha}, {Capozzi}, {Carlsen}, {Carnero-Rosell}, {Carollo},
  {Carrasco-Kind}, {Carretero}, {Castander}, {Clerkin}, {Collett}, {Conselice},
  {Crocce}, {Cunha}, {D'Andrea}, {da Costa}, {Davis}, {Desai}, {Diehl},
  {Dietrich}, {Dodelson}, {Doel}, {Drlica-Wagner}, {Estrada}, {Etherington},
  {Evrard}, {Fabbri}, {Finley}, {Flaugher}, {Foley}, {Fosalba}, {Frieman},
  {Garc{\'\i}a-Bellido}, {Gaztanaga}, {Gerdes}, {Giannantonio}, {Goldstein},
  {Gruen}, {Gruendl}, {Guarnieri}, {Gutierrez}, {Hartley}, {Honscheid}, {Jain},
  {James}, {Jeltema}, {Jouvel}, {Kessler}, {King}, {Kirk}, {Kron}, {Kuehn},
  {Kuropatkin}, {Lahav}, {Li}, {Lima}, {Lin}, {Maia}, {Makler}, {Manera},
  {Maraston}, {Marshall}, {Martini}, {McMahon}, {Melchior}, {Merson}, {Miller},
  {Miquel}, {Mohr}, {Morice-Atkinson}, {Naidoo}, {Neilsen}, {Nichol}, {Nord},
  {Ogando}, {Ostrovski}, {Palmese}, {Papadopoulos}, {Peiris}, {Peoples},
  {Percival}, {Plazas}, {Reed}, {Refregier}, {Romer}, {Roodman}, {Ross},
  {Rozo}, {Rykoff}, {Sadeh}, {Sako}, {S{\'a}nchez}, {Sanchez}, {Santiago},
  {Scarpine}, {Schubnell}, {Sevilla-Noarbe}, {Sheldon}, {Smith}, {Smith},
  {Soares-Santos}, {Sobreira}, {Soumagnac}, {Suchyta}, {Sullivan}, {Swanson},
  {Tarle}, {Thaler}, {Thomas}, {Thomas}, {Tucker}, {Vieira}, {Vikram},
  {Walker}, {Wechsler}, {Weller}, {Wester}, {Whiteway}, {Wilcox}, {Yanny},
  {Zhang}, \& {Zuntz}}]{DES2016}
{Dark Energy Survey Collaboration}, {Abbott}, T., {Abdalla}, F.~B., {et~al.}
  2016, \mnras, 460, 1270

\bibitem[{{de Grijs} \& {Bono}(2015)}]{deGrijs2015}
{de Grijs}, R., \& {Bono}, G. 2015, \aj, 149, 179

\bibitem[{{De Leo} {et~al.}(2020){De Leo}, {Carrera}, {No{\"e}l}, {Read},
  {Erkal}, \& {Gallart}}]{DeLeo2020}
{De Leo}, M., {Carrera}, R., {No{\"e}l}, N. E.~D., {et~al.} 2020, \mnras, 495,
  98

\bibitem[{{Diaz} \& {Bekki}(2012)}]{Diaz2012}
{Diaz}, J.~D., \& {Bekki}, K. 2012, \apj, 750, 36

\bibitem[{{Dobbie} {et~al.}(2014){Dobbie}, {Cole}, {Subramaniam}, \&
  {Keller}}]{Dobbie2014}
{Dobbie}, P.~D., {Cole}, A.~A., {Subramaniam}, A., \& {Keller}, S. 2014,
  \mnras, 442, 1680

\bibitem[{{Drlica-Wagner} {et~al.}(2021){Drlica-Wagner}, {Carlin}, {Nidever},
  {Ferguson}, {Kuropatkin}, {Adam{\'o}w}, {Cerny}, {Choi}, {Esteves},
  {Mart{\'\i}nez-V{\'a}zquez}, {Mau}, {Miller}, {Mutlu-Pakdil}, {Neilsen},
  {Olsen}, {Pace}, {Riley}, {Sakowska}, {Sand}, {Santana-Silva}, {Tollerud},
  {Tucker}, {Vivas}, {Zaborowski}, {Zenteno}, {Abbott}, {Allam}, {Bechtol},
  {Bell}, {Bell}, {Bilaji}, {Bom}, {Carballo-Bello}, {Crnojevi{\'c}}, {Cioni},
  {Diaz-Ocampo}, {de Boer}, {Erkal}, {Gruendl}, {Hernandez-Lang}, {Hughes},
  {James}, {Johnson}, {Li}, {Mao}, {Mart{\'\i}nez-Delgado}, {Massana},
  {McNanna}, {Morgan}, {Nadler}, {No{\"e}l}, {Palmese}, {Peter}, {Rykoff},
  {S{\'a}nchez}, {Shipp}, {Simon}, {Smercina}, {Soares-Santos}, {Stringfellow},
  {Tavangar}, {van der Marel}, {Walker}, {Wechsler}, {Wu}, {Yanny},
  {Fitzpatrick}, {Huang}, {Jacques}, {Nikutta}, {Scott}, \& {Astro Data
  Lab}}]{Delve2021}
{Drlica-Wagner}, A., {Carlin}, J.~L., {Nidever}, D.~L., {et~al.} 2021, \apjs,
  256, 2

\bibitem[{{El Youssoufi} {et~al.}(2021){El Youssoufi}, {Cioni}, {Bell}, {de
  Grijs}, {Groenewegen}, {Ivanov}, {Matijev{\u{i}}c}, {Niederhofer},
  {Oliveira}, {Ripepi}, {Schmidt}, {Subramanian}, {Sun}, \& {van
  Loon}}]{ElYoussoufi2021}
{El Youssoufi}, D., {Cioni}, M.-R.~L., {Bell}, C. P.~M., {et~al.} 2021, \mnras,
  505, 2020

\bibitem[{{Filipovic} {et~al.}(1996){Filipovic}, {White}, {Jones}, {Haynes},
  {Pietsch}, {Wielebinski}, \& {Klein}}]{Filipovic1996}
{Filipovic}, M.~D., {White}, G.~L., {Jones}, P.~A., {et~al.} 1996, in
  Astronomical Society of the Pacific Conference Series, Vol. 112, The History
  of the Milky Way and Its Satellite System, ed. A.~{Burkert}, D.~H.
  {Hartmann}, \& S.~A. {Majewski}, 91

\bibitem[{{Gaia Collaboration} {et~al.}(2021){Gaia Collaboration}, {Luri},
  {Chemin}, {Clementini}, {Delgado}, {McMillan}, {Romero-G{\'o}mez},
  {Balbinot}, {Castro-Ginard}, {Mor}, {Ripepi}, {Sarro}, {Cioni}, {Fabricius},
  {Garofalo}, {Helmi}, {Muraveva}, {Brown}, {Vallenari}, {Prusti}, {de
  Bruijne}, {Babusiaux}, {Biermann}, {Creevey}, {Evans}, {Eyer}, {Hutton},
  {Jansen}, {Jordi}, {Klioner}, {Lammers}, {Lindegren}, {Mignard}, {Panem},
  {Pourbaix}, {Randich}, {Sartoretti}, {Soubiran}, {Walton}, {Arenou},
  {Bailer-Jones}, {Bastian}, {Cropper}, {Drimmel}, {Katz}, {Lattanzi}, {van
  Leeuwen}, {Bakker}, {Casta{\~n}eda}, {De Angeli}, {Ducourant}, {Fouesneau},
  {Fr{\'e}mat}, {Guerra}, {Guerrier}, {Guiraud}, {Jean-Antoine Piccolo},
  {Masana}, {Messineo}, {Mowlavi}, {Nicolas}, {Nienartowicz}, {Pailler},
  {Panuzzo}, {Riclet}, {Roux}, {Seabroke}, {Sordo}, {Tanga}, {Th{\'e}venin},
  {Gracia-Abril}, {Portell}, {Teyssier}, {Altmann}, {Andrae}, {Bellas-Velidis},
  {Benson}, {Berthier}, {Blomme}, {Brugaletta}, {Burgess}, {Busso}, {Carry},
  {Cellino}, {Cheek}, {Damerdji}, {Davidson}, {Delchambre}, {Dell'Oro},
  {Fern{\'a}ndez-Hern{\'a}ndez}, {Galluccio}, {Garc{\'\i}a-Lario},
  {Garcia-Reinaldos}, {Gonz{\'a}lez-N{\'u}{\~n}ez}, {Gosset}, {Haigron},
  {Halbwachs}, {Hambly}, {Harrison}, {Hatzidimitriou}, {Heiter},
  {Hern{\'a}ndez}, {Hestroffer}, {Hodgkin}, {Holl}, {Jan{\ss}en}, {Jevardat de
  Fombelle}, {Jordan}, {Krone-Martins}, {Lanzafame}, {L{\"o}ffler}, {Lorca},
  {Manteiga}, {Marchal}, {Marrese}, {Moitinho}, {Mora}, {Muinonen}, {Osborne},
  {Pancino}, {Pauwels}, {Recio-Blanco}, {Richards}, {Riello}, {Rimoldini},
  {Robin}, {Roegiers}, {Rybizki}, {Siopis}, {Smith}, {Sozzetti}, {Ulla},
  {Utrilla}, {van Leeuwen}, {van Reeven}, {Abbas}, {Abreu Aramburu}, {Accart},
  {Aerts}, {Aguado}, {Ajaj}, {Altavilla}, {{\'A}lvarez}, {{\'A}lvarez
  Cid-Fuentes}, {Alves}, {Anderson}, {Anglada Varela}, {Antoja}, {Audard},
  {Baines}, {Baker}, {Balaguer-N{\'u}{\~n}ez}, {Balog}, {Barache}, {Barbato},
  {Barros}, {Barstow}, {Bartolom{\'e}}, {Bassilana}, {Bauchet},
  {Baudesson-Stella}, {Becciani}, {Bellazzini}, {Bernet}, {Bertone}, {Bianchi},
  {Blanco-Cuaresma}, {Boch}, {Bombrun}, {Bossini}, {Bouquillon}, {Bragaglia},
  {Bramante}, {Breedt}, {Bressan}, {Brouillet}, {Bucciarelli}, {Burlacu},
  {Busonero}, {Butkevich}, {Buzzi}, {Caffau}, {Cancelliere}, {C{\'a}novas},
  {Cantat-Gaudin}, {Carballo}, {Carlucci}, {Carnerero}, {Carrasco},
  {Casamiquela}, {Castellani}, {Castro Sampol}, {Chaoul}, {Charlot},
  {Chiavassa}, {Comoretto}, {Cooper}, {Cornez}, {Cowell}, {Crifo}, {Crosta},
  {Crowley}, {Dafonte}, {Dapergolas}, {David}, {David}, {de Laverny}, {De
  Luise}, {De March}, {De Ridder}, {de Souza}, {de Teodoro}, {de Torres}, {del
  Peloso}, {del Pozo}, {Delgado}, {Delisle}, {Di Matteo}, {Diakite}, {Diener},
  {Distefano}, {Dolding}, {Eappachen}, {Enke}, {Esquej}, {Fabre}, {Fabrizio},
  {Faigler}, {Fedorets}, {Fernique}, {Fienga}, {Figueras}, {Fouron},
  {Fragkoudi}, {Fraile}, {Franke}, {Gai}, {Garabato}, {Garcia-Gutierrez},
  {Garc{\'\i}a-Torres}, {Gavras}, {Gerlach}, {Geyer}, {Giacobbe}, {Gilmore},
  {Girona}, {Giuffrida}, {Gomez}, {Gonzalez-Santamaria}, {Gonz{\'a}lez-Vidal},
  {Granvik}, {Guti{\'e}rrez-S{\'a}nchez}, {Guy}, {Hauser}, {Haywood},
  {Hidalgo}, {Hilger}, {H{\l}adczuk}, {Hobbs}, {Holland}, {Huckle},
  {Jasniewicz}, {Jonker}, {Juaristi Campillo}, {Julbe}, {Karbevska},
  {Kervella}, {Khanna}, {Kochoska}, {Kontizas}, {Kordopatis}, {Korn},
  {Kostrzewa-Rutkowska}, {Kruszy{\'n}ska}, {Lambert}, {Lanza}, {Lasne}, {Le
  Campion}, {Le Fustec}, {Lebreton}, {Lebzelter}, {Leccia}, {Leclerc},
  {Lecoeur-Taibi}, {Liao}, {Licata}, {Lindstr{\o}m}, {Lister}, {Livanou},
  {Lobel}, {Madrero Pardo}, {Managau}, {Mann}, {Marchant}, {Marconi}, {Marcos
  Santos}, {Marinoni}, {Marocco}, {Marshall}, {Martin Polo},
  {Mart{\'\i}n-Fleitas}, {Masip}, {Massari}, {Mastrobuono-Battisti}, {Mazeh},
  {Messina}, {Michalik}, {Millar}, {Mints}, {Molina}, {Molinaro}, {Moln{\'a}r},
  {Montegriffo}, {Morbidelli}, {Morel}, {Morris}, {Mulone}, {Munoz}, {Murphy},
  {Musella}, {Noval}, {Ord{\'e}novic}, {Orr{\`u}}, {Osinde}, {Pagani},
  {Pagano}, {Palaversa}, {Palicio}, {Panahi}, {Pawlak}, {Pe{\~n}alosa
  Esteller}, {Penttil{\"a}}, {Piersimoni}, {Pineau}, {Plachy}, {Plum},
  {Poggio}, {Poretti}, {Poujoulet}, {Pr{\v{s}}a}, {Pulone}, {Racero},
  {Ragaini}, {Rainer}, {Raiteri}, {Rambaux}, {Ramos}, {Ramos-Lerate}, {Re
  Fiorentin}, {Regibo}, {Reyl{\'e}}, {Riva}, {Rixon}, {Robichon}, {Robin},
  {Roelens}, {Rohrbasser}, {Rowell}, {Royer}, {Rybicki}, {Sadowski},
  {Sagrist{\`a} Sell{\'e}s}, {Sahlmann}, {Salgado}, {Salguero}, {Samaras},
  {Gimenez}, {Sanna}, {Santove{\~n}a}, {Sarasso}, {Schultheis}, {Sciacca},
  {Segol}, {Segovia}, {S{\'e}gransan}, {Semeux}, {Siddiqui}, {Siebert},
  {Siltala}, {Slezak}, {Smart}, {Solano}, {Solitro}, {Souami}, {Souchay},
  {Spagna}, {Spoto}, {Steele}, {Steidelm{\"u}ller}, {Stephenson},
  {S{\"u}veges}, {Szabados}, {Szegedi-Elek}, {Taris}, {Tauran}, {Taylor},
  {Teixeira}, {Thuillot}, {Tonello}, {Torra}, {Torra}, {Turon}, {Unger},
  {Vaillant}, {van Dillen}, {Vanel}, {Vecchiato}, {Viala}, {Vicente},
  {Voutsinas}, {Weiler}, {Wevers}, {Wyrzykowski}, {Yoldas}, {Yvard}, {Zhao},
  {Zorec}, {Zucker}, {Zurbach}, \& {Zwitter}}]{Gaia2021}
{Gaia Collaboration}, {Luri}, X., {Chemin}, L., {et~al.} 2021, \aap, 649, A7

\bibitem[{{Garc{\'{\i}}a P{\'e}rez} {et~al.}(2016){Garc{\'{\i}}a P{\'e}rez},
  {Allende Prieto}, {Holtzman}, {Shetrone}, {M{\'e}sz{\'a}ros}, {Bizyaev},
  {Carrera}, {Cunha}, {Garc{\'{\i}}a-Hern{\'a}ndez}, {Johnson}, {Majewski},
  {Nidever}, {Schiavon}, {Shane}, {Smith}, {Sobeck}, {Troup}, {Zamora},
  {Weinberg}, {Bovy}, {Eisenstein}, {Feuillet}, {Frinchaboy}, {Hayden},
  {Hearty}, {Nguyen}, {O{\'}Connell}, {Pinsonneault}, {Wilson}, \&
  {Zasowski}}]{Garcia2016}
{Garc{\'{\i}}a P{\'e}rez}, A.~E., {Allende Prieto}, C., {Holtzman}, J.~A.,
  {et~al.} 2016, \aj, 151, 144

\bibitem[{{Gardiner} \& {Hawkins}(1991)}]{Gardiner1991}
{Gardiner}, L.~T., \& {Hawkins}, M.~R.~S. 1991, \mnras, 251, 174

\bibitem[{{Gardiner} \& {Noguchi}(1996)}]{Gardiner1996}
{Gardiner}, L.~T., \& {Noguchi}, M. 1996, \mnras, 278, 191

\bibitem[{{Groenewegen} {et~al.}(2019){Groenewegen}, {Cioni}, {Girardi}, {de
  Grijs}, {Ivanov}, {Marconi}, {Muraveva}, {Ripepi}, \& {van
  Loon}}]{Groenewegen2019}
{Groenewegen}, M.~A.~T., {Cioni}, M. R.~L., {Girardi}, L., {et~al.} 2019, \aap,
  622, A63

\bibitem[{{Gunn} {et~al.}(2006){Gunn}, {Siegmund}, {Mannery}, {Owen}, {Hull},
  {Leger}, {Carey}, {Knapp}, {York}, {Boroski}, {Kent}, {Lupton}, {Rockosi},
  {Evans}, {Waddell}, {Anderson}, {Annis}, {Barentine}, {Bartoszek}, {Bastian},
  {Bracker}, {Brewington}, {Briegel}, {Brinkmann}, {Brown}, {Carr},
  {Czarapata}, {Drennan}, {Dombeck}, {Federwitz}, {Gillespie}, {Gonzales},
  {Hansen}, {Harvanek}, {Hayes}, {Jordan}, {Kinney}, {Klaene}, {Kleinman},
  {Kron}, {Kresinski}, {Lee}, {Limmongkol}, {Lindenmeyer}, {Long}, {Loomis},
  {McGehee}, {Mantsch}, {Neilsen}, {Neswold}, {Newman}, {Nitta}, {Peoples},
  {Pier}, {Prieto}, {Prosapio}, {Rivetta}, {Schneider}, {Snedden}, \&
  {Wang}}]{Gunn2006}
{Gunn}, J.~E., {Siegmund}, W.~A., {Mannery}, E.~J., {et~al.} 2006, \aj, 131,
  2332

\bibitem[{{Gustafsson} {et~al.}(2008){Gustafsson}, {Edvardsson}, {Eriksson},
  {J{\o}rgensen}, {Nordlund}, \& {Plez}}]{Gustafsson2008}
{Gustafsson}, B., {Edvardsson}, B., {Eriksson}, K., {et~al.} 2008, \aap, 486,
  951

\bibitem[{{Harris} \& {Zaritsky}(2004)}]{harriszaritsky2004}
{Harris}, J., \& {Zaritsky}, D. 2004, \aj, 127, 1531

\bibitem[{{Haschke} {et~al.}(2012){Haschke}, {Grebel}, \&
  {Duffau}}]{Haschke2012}
{Haschke}, R., {Grebel}, E.~K., \& {Duffau}, S. 2012, \aj, 144, 107

\bibitem[{{Hasselquist} {et~al.}(2016){Hasselquist}, {Shetrone}, {Cunha},
  {Smith}, {Holtzman}, {Lawler}, {Allende Prieto}, {Beers}, {Chojnowski},
  {Fern{\'a}ndez-Trincado}, {Garc{\'{\i}}a-Hern{\'a}ndez}, {Hearty},
  {Majewski}, {Pereira}, {Placco}, {Villanova}, \& {Zamora}}]{Hasselquist2016}
{Hasselquist}, S., {Shetrone}, M., {Cunha}, K., {et~al.} 2016, \apj, 833, 81

\bibitem[{{Hatzidimitriou} {et~al.}(1993){Hatzidimitriou}, {Cannon}, \&
  {Hawkins}}]{1993MNRAS.261..873H}
{Hatzidimitriou}, D., {Cannon}, R.~D., \& {Hawkins}, M.~R.~S. 1993, \mnras,
  261, 873

\bibitem[{{Hatzidimitriou} \& {Hawkins}(1989)}]{Hatzidimitriou1989}
{Hatzidimitriou}, D., \& {Hawkins}, M.~R.~S. 1989, \mnras, 241, 667

\bibitem[{{Hindman} {et~al.}(1961){Hindman}, {McGee}, {Carter}, \&
  {Kerr}}]{Hindman1961}
{Hindman}, J.~V., {McGee}, R.~X., {Carter}, A.~W.~L., \& {Kerr}, F.~J. 1961,
  \aj, 66, 45

\bibitem[{{Hubeny} \& {Lanz}(2011)}]{Hubeny2011}
{Hubeny}, I., \& {Lanz}, T. 2011, {Synspec: General Spectrum Synthesis
  Program}, Astrophysics Source Code Library, record ascl:1109.022, , ,
  ascl:1109.022

\bibitem[{{Irrgang} {et~al.}(2013){Irrgang}, {Wilcox}, {Tucker}, \&
  {Schiefelbein}}]{Irrgang2013}
{Irrgang}, A., {Wilcox}, B., {Tucker}, E., \& {Schiefelbein}, L. 2013, \aap,
  549, A137

\bibitem[{{Jacyszyn-Dobrzeniecka} {et~al.}(2017){Jacyszyn-Dobrzeniecka},
  {Skowron}, {Mr{\'o}z}, {Soszy{\'n}ski}, {Udalski}, {Pietrukowicz}, {Skowron},
  {Poleski}, {Koz{\l}owski}, {Wyrzykowski}, {Pawlak}, {Szyma{\'n}ski}, \&
  {Ulaczyk}}]{Jacyszyn-Dobrzeniecka2017}
{Jacyszyn-Dobrzeniecka}, A.~M., {Skowron}, D.~M., {Mr{\'o}z}, P., {et~al.}
  2017, \actaa, 67, 1

\bibitem[{{Jacyszyn-Dobrzeniecka} {et~al.}(2020){Jacyszyn-Dobrzeniecka},
  {Mr{\'o}z}, {Kruszy{\'n}ska}, {Soszy{\'n}ski}, {Skowron}, {Udalski},
  {Szyma{\'n}ski}, {Iwanek}, {Skowron}, {Pietrukowicz}, {Poleski},
  {Koz{\l}owski}, {Ulaczyk}, {Rybicki}, \& {Wrona}}]{Jacyszyn2020}
{Jacyszyn-Dobrzeniecka}, A.~M., {Mr{\'o}z}, P., {Kruszy{\'n}ska}, K., {et~al.}
  2020, \apj, 889, 26

\bibitem[{{James} {et~al.}(2021){James}, {Subramanian}, {Omkumar}, {Mary},
  {Bekki}, {Cioni}, {de Grijs}, {El Youssoufi}, {Kartha}, {Niederhofer}, \&
  {van Loon}}]{James2021}
{James}, D., {Subramanian}, S., {Omkumar}, A.~O., {et~al.} 2021, \mnras, 508,
  5854

\bibitem[{{J{\"o}nsson} {et~al.}(2020){J{\"o}nsson}, {Holtzman}, {Allende
  Prieto}, {Cunha}, {Garc{\'\i}a-Hern{\'a}ndez}, {Hasselquist}, {Masseron},
  {Osorio}, {Shetrone}, {Smith}, {Stringfellow}, {Bizyaev}, {Edvardsson},
  {Majewski}, {M{\'e}sz{\'a}ros}, {Souto}, {Zamora}, {Beaton}, {Bovy}, {Donor},
  {Pinsonneault}, {Poovelil}, \& {Sobeck}}]{Jonsson2020}
{J{\"o}nsson}, H., {Holtzman}, J.~A., {Allende Prieto}, C., {et~al.} 2020, \aj,
  160, 120

\bibitem[{{Kallivayalil} {et~al.}(2013){Kallivayalil}, {van der Marel},
  {Besla}, {Anderson}, \& {Alcock}}]{Kallivayalil2013}
{Kallivayalil}, N., {van der Marel}, R.~P., {Besla}, G., {Anderson}, J., \&
  {Alcock}, C. 2013, \apj, 764, 161

\bibitem[{{Laney} \& {Stobie}(1986)}]{Laney1986}
{Laney}, C.~D., \& {Stobie}, R.~S. 1986, \mnras, 222, 449

\bibitem[{{Leung} \& {Bovy}(2019)}]{2019MNRAS.489.2079L}
{Leung}, H.~W., \& {Bovy}, J. 2019, \mnras, 489, 2079

\bibitem[{{Luyten}(1922)}]{Luyten1922}
{Luyten}, W.~J. 1922, Lick Observatory Bulletin, 336, 135

\bibitem[{{Majewski} {et~al.}(2017){Majewski}, {Schiavon}, {Frinchaboy},
  {Allende Prieto}, {Barkhouser}, {Bizyaev}, {Blank}, {Brunner}, {Burton},
  {Carrera}, {Chojnowski}, {Cunha}, {Epstein}, {Fitzgerald}, {Garc{\'{\i}}a
  P{\'e}rez}, {Hearty}, {Henderson}, {Holtzman}, {Johnson}, {Lam}, {Lawler},
  {Maseman}, {M{\'e}sz{\'a}ros}, {Nelson}, {Nguyen}, {Nidever}, {Pinsonneault},
  {Shetrone}, {Smee}, {Smith}, {Stolberg}, {Skrutskie}, {Walker}, {Wilson},
  {Zasowski}, {Anders}, {Basu}, {Beland}, {Blanton}, {Bovy}, {Brownstein},
  {Carlberg}, {Chaplin}, {Chiappini}, {Eisenstein}, {Elsworth}, {Feuillet},
  {Fleming}, {Galbraith-Frew}, {Garc{\'{\i}}a}, {Garc{\'{\i}}a-Hern{\'a}ndez},
  {Gillespie}, {Girardi}, {Gunn}, {Hasselquist}, {Hayden}, {Hekker}, {Ivans},
  {Kinemuchi}, {Klaene}, {Mahadevan}, {Mathur}, {Mosser}, {Muna}, {Munn},
  {Nichol}, {O'Connell}, {Parejko}, {Robin}, {Rocha-Pinto}, {Schultheis},
  {Serenelli}, {Shane}, {Silva Aguirre}, {Sobeck}, {Thompson}, {Troup},
  {Weinberg}, \& {Zamora}}]{Majewski2017}
{Majewski}, S.~R., {Schiavon}, R.~P., {Frinchaboy}, P.~M., {et~al.} 2017, \aj,
  154, 94

\bibitem[{{Massana} {et~al.}(2020){Massana}, {No{\"e}l}, {Nidever}, {Erkal},
  {de Boer}, {Choi}, {Majewski}, {Olsen}, {Monachesi}, {Gallart}, {Marel},
  {Ruiz-Lara}, {Zaritsky}, {Martin}, {Mu{\~n}oz}, {Cioni}, {Bell}, {Bell},
  {Stringfellow}, {Belokurov}, {Monelli}, {Walker}, {Mart{\'\i}nez-Delgado},
  {Vivas}, \& {Conn}}]{Massana2020}
{Massana}, P., {No{\"e}l}, N. E.~D., {Nidever}, D.~L., {et~al.} 2020, \mnras,
  498, 1034

\bibitem[{{Massana} {et~al.}(2022){Massana}, {Ruiz-Lara}, {No{\"e}l},
  {Gallart}, {Nidever}, {Choi}, {Sakowska}, {Besla}, {Olsen}, {Monelli},
  {Dorta}, {Stringfellow}, {Cassisi}, {Bernard}, {Zaritsky}, {Cioni},
  {Monachesi}, {van der Marel}, {de Boer}, \& {Walker}}]{massana2022}
{Massana}, P., {Ruiz-Lara}, T., {No{\"e}l}, N.~E.~D., {et~al.} 2022, \mnras,
  513, L40

\bibitem[{{Mu{\~n}oz} {et~al.}(2023){Mu{\~n}oz}, {Monachesi}, {Nidever},
  {Majewski}, {Cheng}, {Olsen}, {Choi}, {Zivick}, {Geisler}, {Almeida},
  {Mu{\~n}oz}, {Nitschelm}, {Roman-Lopes}, {Lane}, \&
  {Fern{\'a}ndez-Trincado}}]{munoz2023}
{Mu{\~n}oz}, C., {Monachesi}, A., {Nidever}, D.~L., {et~al.} 2023, arXiv
  e-prints, arXiv:2305.19460

\bibitem[{{Muller} \& {Bekki}(2007)}]{Muller2007}
{Muller}, E., \& {Bekki}, K. 2007, \mnras, 381, L11

\bibitem[{{Muller} {et~al.}(2003){Muller}, {Staveley-Smith}, {Zealey}, \&
  {Stanimirovi{\'c}}}]{Muller2003}
{Muller}, E., {Staveley-Smith}, L., {Zealey}, W., \& {Stanimirovi{\'c}}, S.
  2003, \mnras, 339, 105

\bibitem[{{Nidever}(2021)}]{Doppler2021}
{Nidever}, D. 2021, {dnidever/doppler: Cannon and Payne models}, Zenodo,
  vv1.1.0,  Zenodo, doi:10.5281/zenodo.4906680

\bibitem[{{Nidever} {et~al.}(2008){Nidever}, {Majewski}, \& {Butler
  Burton}}]{Nidever2008}
{Nidever}, D.~L., {Majewski}, S.~R., \& {Butler Burton}, W. 2008, \apj, 679,
  432

\bibitem[{{Nidever} {et~al.}(2013){Nidever}, {Monachesi}, {Bell}, {Majewski},
  {Mu{\~n}oz}, \& {Beaton}}]{Nidever2013}
{Nidever}, D.~L., {Monachesi}, A., {Bell}, E.~F., {et~al.} 2013, \apj, 779, 145

\bibitem[{{Nidever} {et~al.}(2015{\natexlab{a}}){Nidever}, {Holtzman}, {Allende
  Prieto}, {Beland}, {Bender}, {Bizyaev}, {Burton}, {Desphande}, {Fleming},
  {Garc{\'{\i}}a P{\'e}rez}, {Hearty}, {Majewski}, {M{\'e}sz{\'a}ros}, {Muna},
  {Nguyen}, {Schiavon}, {Shetrone}, {Skrutskie}, {Sobeck}, \&
  {Wilson}}]{Nidever2015}
{Nidever}, D.~L., {Holtzman}, J.~A., {Allende Prieto}, C., {et~al.}
  2015{\natexlab{a}}, \aj, 150, 173

\bibitem[{{Nidever} {et~al.}(2015{\natexlab{b}}){Nidever}, {Holtzman}, {Allende
  Prieto}, {Beland}, {Bender}, {Bizyaev}, {Burton}, {Desphande}, {Fleming},
  {Garc{\'{\i}}a P{\'e}rez}, {Hearty}, {Majewski}, {M{\'e}sz{\'a}ros}, {Muna},
  {Nguyen}, {Schiavon}, {Shetrone}, {Skrutskie}, {Sobeck}, \& {Wilson}}]{dln15}
---. 2015{\natexlab{b}}, \aj, 150, 173

\bibitem[{{Nidever} {et~al.}(2017){Nidever}, {Olsen}, {Walker}, {Vivas},
  {Blum}, {Kaleida}, {Choi}, {Conn}, {Gruendl}, {Bell}, {Besla}, {Mu{\~n}oz},
  {Gallart}, {Martin}, {Olszewski}, {Saha}, {Monachesi}, {Monelli}, {de Boer},
  {Johnson}, {Zaritsky}, {Stringfellow}, {van der Marel}, {Cioni}, {Jin},
  {Majewski}, {Martinez-Delgado}, {Monteagudo}, {No{\"e}l}, {Bernard},
  {Kunder}, {Chu}, {Bell}, {Santana}, {Frechem}, {Medina}, {Parkash},
  {Navarrete}, \& {Hayes}}]{Smash2017}
{Nidever}, D.~L., {Olsen}, K., {Walker}, A.~R., {et~al.} 2017, \aj, 154, 199

\bibitem[{{Nidever} {et~al.}(2019){Nidever}, {Olsen}, {Choi}, {de Boer},
  {Blum}, {Bell}, {Zaritsky}, {Martin}, {Saha}, {Conn}, {Besla}, {van der
  Marel}, {No{\"e}l}, {Monachesi}, {Stringfellow}, {Massana}, {Cioni},
  {Gallart}, {Monelli}, {Martinez-Delgado}, {Mu{\~n}oz}, {Majewski}, {Vivas},
  {Walker}, {Kaleida}, \& {Chu}}]{Nidever2019}
{Nidever}, D.~L., {Olsen}, K., {Choi}, Y., {et~al.} 2019, \apj, 874, 118

\bibitem[{{Nidever} {et~al.}(2020{\natexlab{a}}){Nidever}, {Hasselquist},
  {Hayes}, {Hawkins}, {Povick}, {Majewski}, {Smith}, {Anguiano},
  {Stringfellow}, {Sobeck}, {Cunha}, {Beers}, {Bestenlehner}, {Cohen},
  {Garcia-Hernandez}, {J{\"o}nsson}, {Nitschelm}, {Shetrone}, {Lacerna},
  {Allende Prieto}, {Beaton}, {Dell'Agli}, {Fern{\'a}ndez-Trincado},
  {Feuillet}, {Gallart}, {Hearty}, {Holtzman}, {Manchado}, {Mu{\~n}oz},
  {O'Connell}, \& {Rosado}}]{Nidever2020}
{Nidever}, D.~L., {Hasselquist}, S., {Hayes}, C.~R., {et~al.}
  2020{\natexlab{a}}, \apj, 895, 88

\bibitem[{{Nidever} {et~al.}(2020{\natexlab{b}}){Nidever}, {Hasselquist},
  {Hayes}, {Hawkins}, {Povick}, {Majewski}, {Smith}, {Anguiano},
  {Stringfellow}, {Sobeck}, {Cunha}, {Beers}, {Bestenlehner}, {Cohen},
  {Garcia-Hernandez}, {J{\"o}nsson}, {Nitschelm}, {Shetrone}, {Lacerna},
  {Allende Prieto}, {Beaton}, {Dell'Agli}, {Fern{\'a}ndez-Trincado},
  {Feuillet}, {Gallart}, {Hearty}, {Holtzman}, {Manchado}, {Mu{\~n}oz},
  {O'Connell}, \& {Rosado}}]{2020ApJ...895...88N}
---. 2020{\natexlab{b}}, \apj, 895

\bibitem[{{Omkumar} {et~al.}(2021){Omkumar}, {Subramanian}, {Niederhofer},
  {Diaz}, {Cioni}, {El Youssoufi}, {Bekki}, {de Grijs}, \& {van
  Loon}}]{Omkumar2021}
{Omkumar}, A.~O., {Subramanian}, S., {Niederhofer}, F., {et~al.} 2021, \mnras,
  500, 2757

\bibitem[{{Osorio} {et~al.}(2020){Osorio}, {Allende Prieto}, {Hubeny},
  {M{\'e}sz{\'a}ros}, \& {Shetrone}}]{Osorio2020}
{Osorio}, Y., {Allende Prieto}, C., {Hubeny}, I., {M{\'e}sz{\'a}ros}, S., \&
  {Shetrone}, M. 2020, \aap, 637, A80

\bibitem[{{Pieres} {et~al.}(2017){Pieres}, {Santiago}, {Drlica-Wagner},
  {Bechtol}, {Marel}, {Besla}, {Martin}, {Belokurov}, {Gallart},
  {Martinez-Delgado}, {Marshall}, {N{\"o}el}, {Majewski}, {Cioni}, {Li},
  {Hartley}, {Luque}, {Conn}, {Walker}, {Balbinot}, {Stringfellow}, {Olsen},
  {Nidever}, {da Costa}, {Ogando}, {Maia}, {Neto}, {Abbott}, {Abdalla},
  {Allam}, {Annis}, {Benoit-L{\'e}vy}, {Rosell}, {Kind}, {Carretero}, {Cunha},
  {D'Andrea}, {Desai}, {Diehl}, {Doel}, {Flaugher}, {Fosalba},
  {Garc{\'\i}a-Bellido}, {Gruen}, {Gruendl}, {Gschwend}, {Gutierrez},
  {Honscheid}, {James}, {Kuehn}, {Kuropatkin}, {Menanteau}, {Miquel}, {Plazas},
  {Romer}, {Sako}, {Sanchez}, {Scarpine}, {Schubnell}, {Sevilla-Noarbe},
  {Smith}, {Soares-Santos}, {Sobreira}, {Suchyta}, {Swanson}, {Tarle},
  {Tucker}, \& {Wester}}]{Pieres2017}
{Pieres}, A., {Santiago}, B.~X., {Drlica-Wagner}, A., {et~al.} 2017, \mnras,
  468, 1349

\bibitem[{{Putman} {et~al.}(2003){Putman}, {Staveley-Smith}, {Freeman},
  {Gibson}, \& {Barnes}}]{Putman2003}
{Putman}, M.~E., {Staveley-Smith}, L., {Freeman}, K.~C., {Gibson}, B.~K., \&
  {Barnes}, D.~G. 2003, \apj, 586, 170

\bibitem[{{Putman} {et~al.}(1998){Putman}, {Gibson}, {Staveley-Smith}, {Banks},
  {Barnes}, {Bhatal}, {Disney}, {Ekers}, {Freeman}, {Haynes}, {Henning},
  {Jerjen}, {Kilborn}, {Koribalski}, {Knezek}, {Malin}, {Mould}, {Oosterloo},
  {Price}, {Ryder}, {Sadler}, {Stewart}, {Stootman}, {Vaile}, {Webster}, \&
  {Wright}}]{Putman1998}
{Putman}, M.~E., {Gibson}, B.~K., {Staveley-Smith}, L., {et~al.} 1998, \nat,
  394, 752

\bibitem[{{Queiroz} {et~al.}(2020){Queiroz}, {Anders}, {Chiappini},
  {Khalatyan}, {Santiago}, {Steinmetz}, {Valentini}, {Miglio}, {Bossini},
  {Barbuy}, {Minchev}, {Minniti}, {Garc{\'\i}a Hern{\'a}ndez}, {Schultheis},
  {Beaton}, {Beers}, {Bizyaev}, {Brownstein}, {Cunha},
  {Fern{\'a}ndez-Trincado}, {Frinchaboy}, {Lane}, {Majewski}, {Nataf},
  {Nitschelm}, {Pan}, {Roman-Lopes}, {Sobeck}, {Stringfellow}, \&
  {Zamora}}]{2020A&A...638A..76Q}
{Queiroz}, A.~B.~A., {Anders}, F., {Chiappini}, C., {et~al.} 2020, \aap, 638,
  A76

\bibitem[{{Rubele} {et~al.}(2018){Rubele}, {Pastorelli}, {Girardi}, {Cioni},
  {Zaggia}, {Marigo}, {Bekki}, {Bressan}, {Clementini}, {de Grijs}, {Emerson},
  {Groenewegen}, {Ivanov}, {Muraveva}, {Nanni}, {Oliveira}, {Ripepi}, {Sun}, \&
  {van Loon}}]{rubele2018}
{Rubele}, S., {Pastorelli}, G., {Girardi}, L., {et~al.} 2018, \mnras, 478, 5017

\bibitem[{{Santana} {et~al.}(2021){Santana}, {Beaton}, {Covey}, {O'Connell},
  {Longa-Pe{\~n}a}, {Cohen}, {Fern{\'a}ndez-Trincado}, {Hayes}, {Zasowski},
  {Sobeck}, {Majewski}, {Chojnowski}, {De Lee}, {Oelkers}, {Stringfellow},
  {Almeida}, {Anguiano}, {Donor}, {Frinchaboy}, {Hasselquist}, {Johnson},
  {Kollmeier}, {Nidever}, {Price-Whelan}, {Rojas-Arriagada}, {Schultheis},
  {Shetrone}, {Simon}, {Aerts}, {Borissova}, {Drout}, {Geisler}, {Law},
  {Medina}, {Minniti}, {Monachesi}, {Mu{\~n}oz}, {Poleski}, {Roman-Lopes},
  {Schlaufman}, {Stutz}, {Teske}, {Tkachenko}, {Van Saders}, {Weinberger}, \&
  {Zoccali}}]{Santana2021}
{Santana}, F.~A., {Beaton}, R.~L., {Covey}, K.~R., {et~al.} 2021, \aj, 162, 303

\bibitem[{{Shapley}(1940)}]{Shapley1940}
{Shapley}, H. 1940, Harvard College Observatory Bulletin, 914, 8

\bibitem[{{Shetrone} {et~al.}(2015){Shetrone}, {Bizyaev}, {Lawler}, {Allende
  Prieto}, {Johnson}, {Smith}, {Cunha}, {Holtzman}, {Garc{\'{\i}}a P{\'e}rez},
  {M{\'e}sz{\'a}ros}, {Sobeck}, {Zamora}, {Garc{\'{\i}}a-Hern{\'a}ndez},
  {Souto}, {Chojnowski}, {Koesterke}, {Majewski}, \& {Zasowski}}]{shetrone2015}
{Shetrone}, M., {Bizyaev}, D., {Lawler}, J.~E., {et~al.} 2015, \apjs, 221, 24

\bibitem[{{Smith} {et~al.}(2021){Smith}, {Bizyaev}, {Cunha}, {Shetrone},
  {Souto}, {Allende Prieto}, {Masseron}, {M{\'e}sz{\'a}ros}, {J{\"o}nsson},
  {Hasselquist}, {Osorio}, {Garc{\'\i}a-Hern{\'a}ndez}, {Plez}, {Beaton},
  {Holtzman}, {Majewski}, {Stringfellow}, \& {Sobeck}}]{Smith2021}
{Smith}, V.~V., {Bizyaev}, D., {Cunha}, K., {et~al.} 2021, \aj, 161, 254

\bibitem[{{Subramanian} {et~al.}(2017){Subramanian}, {Rubele}, {Sun},
  {Girardi}, {de Grijs}, {van Loon}, {Cioni}, {Piatti}, {Bekki}, {Emerson},
  {Ivanov}, {Kerber}, {Marconi}, {Ripepi}, \& {Tatton}}]{Subramanian2017}
{Subramanian}, S., {Rubele}, S., {Sun}, N.-C., {et~al.} 2017, \mnras, 467, 2980

\bibitem[{{Tatton} {et~al.}(2021){Tatton}, {van Loon}, {Cioni}, {Bekki},
  {Bell}, {Choudhury}, {de Grijs}, {Groenewegen}, {Ivanov}, {Marconi},
  {Oliveira}, {Ripepi}, {Rubele}, {Subramanian}, \& {Sun}}]{Tatton2021}
{Tatton}, B.~L., {van Loon}, J.~T., {Cioni}, M. R.~L., {et~al.} 2021, \mnras,
  504, 2983

\bibitem[{{van der Marel} {et~al.}(2002){van der Marel}, {Alves}, {Hardy}, \&
  {Suntzeff}}]{vdm2002}
{van der Marel}, R.~P., {Alves}, D.~R., {Hardy}, E., \& {Suntzeff}, N.~B. 2002,
  \aj, 124, 2639

\bibitem[{{Westerlund} \& {Glaspey}(1971)}]{Westerlund1971}
{Westerlund}, B.~E., \& {Glaspey}, J. 1971, \aap, 10, 1

\bibitem[{{Wilson} {et~al.}(2019){Wilson}, {Hearty}, {Skrutskie}, {Majewski},
  {Holtzman}, {Eisenstein}, {Gunn}, {Blank}, {Henderson}, {Smee}, {Nelson},
  {Nidever}, {Arns}, {Barkhouser}, {Barr}, {Beland}, {Bershady}, {Blanton},
  {Brunner}, {Burton}, {Carey}, {Carr}, {Colque}, {Crane}, {Damke}, {Davidson},
  {Dean}, {Di Mille}, {Don}, {Ebelke}, {Evans}, {Fitzgerald}, {Gillespie},
  {Hall}, {Harding}, {Harding}, {Hammond}, {Hancock}, {Harrison}, {Hope},
  {Horne}, {Karakla}, {Lam}, {Leger}, {MacDonald}, {Maseman}, {Matsunari},
  {Melton}, {Mitcheltree}, {O'Brien}, {O'Connell}, {Patten}, {Richardson},
  {Rieke}, {Rieke}, {Roman-Lopes}, {Schiavon}, {Sobeck}, {Stolberg}, {Stoll},
  {Tembe}, {Trujillo}, {Uomoto}, {Vernieri}, {Walker}, {Weinberg}, {Young},
  {Anthony-Brumfield}, {Bizyaev}, {Breslauer}, {De Lee}, {Downey}, {Halverson},
  {Huehnerhoff}, {Klaene}, {Leon}, {Long}, {Mahadevan}, {Malanushenko},
  {Nguyen}, {Owen}, {S{\'a}nchez-Gallego}, {Sayres}, {Shane}, {Shectman},
  {Shetrone}, {Skinner}, {Stauffer}, \& {Zhao}}]{Wilson2019}
{Wilson}, J.~C., {Hearty}, F.~R., {Skrutskie}, M.~F., {et~al.} 2019, \pasp,
  131, 055001

\bibitem[{{Zamora} {et~al.}(2015){Zamora}, {Garc{\'{\i}}a-Hern{\'a}ndez},
  {Allende Prieto}, {Carrera}, {Koesterke}, {Edvardsson}, {Castelli}, {Plez},
  {Bizyaev}, {Cunha}, {Garc{\'{\i}}a P{\'e}rez}, {Gustafsson}, {Holtzman},
  {Lawler}, {Majewski}, {Manchado}, {M{\'e}sz{\'a}ros}, {Shane}, {Shetrone},
  {Smith}, \& {Zasowski}}]{zamora2015}
{Zamora}, O., {Garc{\'{\i}}a-Hern{\'a}ndez}, D.~A., {Allende Prieto}, C.,
  {et~al.} 2015, \aj, 149, 181

\bibitem[{{Zasowski} {et~al.}(2013){Zasowski}, {Johnson}, {Frinchaboy},
  {Majewski}, {Nidever}, {Rocha Pinto}, {Girardi}, {Andrews}, {Chojnowski},
  {Cudworth}, {Jackson}, {Munn}, {Skrutskie}, {Beaton}, {Blake}, {Covey},
  {Deshpande}, {Epstein}, {Fabbian}, {Fleming}, {Garcia Hernandez}, {Herrero},
  {Mahadevan}, {M{\'e}sz{\'a}ros}, {Schultheis}, {Sellgren}, {Terrien}, {van
  Saders}, {Allende Prieto}, {Bizyaev}, {Burton}, {Cunha}, {da Costa},
  {Hasselquist}, {Hearty}, {Holtzman}, {Garc{\'\i}a P{\'e}rez}, {Maia},
  {O'Connell}, {O'Donnell}, {Pinsonneault}, {Santiago}, {Schiavon}, {Shetrone},
  {Smith}, \& {Wilson}}]{zas13}
{Zasowski}, G., {Johnson}, J.~A., {Frinchaboy}, P.~M., {et~al.} 2013, \aj, 146,
  81

\bibitem[{{Zasowski} {et~al.}(2017){Zasowski}, {Cohen}, {Chojnowski},
  {Santana}, {Oelkers}, {Andrews}, {Beaton}, {Bender}, {Bird}, {Bovy},
  {Carlberg}, {Covey}, {Cunha}, {Dell'Agli}, {Fleming}, {Frinchaboy},
  {Garc{\'{\i}}a-Hern{\'a}ndez}, {Harding}, {Holtzman}, {Johnson}, {Kollmeier},
  {Majewski}, {M{\'e}sz{\'a}ros}, {Munn}, {Mu{\~n}oz}, {Ness}, {Nidever},
  {Poleski}, {Rom{\'a}n-Z{\'u}{\~n}iga}, {Shetrone}, {Simon}, {Smith},
  {Sobeck}, {Stringfellow}, {Szigeti{\'a}ros}, {Tayar}, \& {Troup}}]{zas17}
{Zasowski}, G., {Cohen}, R.~E., {Chojnowski}, S.~D., {et~al.} 2017, \aj, 154,
  198

\bibitem[{{Zivick} {et~al.}(2021){Zivick}, {Kallivayalil}, \& {van der
  Marel}}]{Zivick2021}
{Zivick}, P., {Kallivayalil}, N., \& {van der Marel}, R.~P. 2021, \apj, 910, 36

\bibitem[{{Zivick} {et~al.}(2018){Zivick}, {Kallivayalil}, {van der Marel},
  {Besla}, {Linden}, {Koz{\l}owski}, {Fritz}, {Kochanek}, {Anderson}, {Sohn},
  {Geha}, \& {Alcock}}]{Zivick2018}
{Zivick}, P., {Kallivayalil}, N., {van der Marel}, R.~P., {et~al.} 2018, \apj,
  864, 55

\bibitem[{{Zivick} {et~al.}(2019){Zivick}, {Kallivayalil}, {Besla}, {Sohn},
  {van der Marel}, {del Pino}, {Linden}, {Fritz}, \& {Anderson}}]{Zivick2019}
{Zivick}, P., {Kallivayalil}, N., {Besla}, G., {et~al.} 2019, \apj, 874, 78

\end{thebibliography}


\end{document}